\documentclass[a4paper,twocolumn,10pt]{article}
\usepackage[pdftex]{graphicx, color}
\usepackage{authblk}

\leftmargin  \leftmargini
\def\abstract{\topsep=0pt\partopsep=0pt\parsep=0pt\itemsep=0pt\relax
\trivlist\item[\hskip\labelsep
{\bfseries\abstractname}]\if!\abstractname!\hskip-\labelsep\fi}
    
\begin{document}

\title{Aspects of weakening in granular assemblies}
\author{Philipp Welker}
\affil{
Institut f\"ur Computerphysik,\\
Universit\"at Stuttgart,\\
70569 Stuttgart, Germany}
 \author{Sean McNamara}
\affil{
Institut de Physique de Rennes,\\
Universit\'e de Rennes 1,\\
35042 Rennes cedex, France
}

\date{(Created: \today)}

\maketitle

\begin{abstract}
We study numerical simulations of large ($N\approx 10^4$) two-dimensional quasi-static granular assemblies subjected to a slowly increasing deviator stress.
We report some peculiarities in the behavior of these packings that have not yet been adressed. The number of sliding contacts is not necessarily related to stability: first the number of sliding contacts rises linearly and smoothly with the applied stress.
Then, at approximately half the peak stress, the increase slows down, a plateau develops, and a decrease follows.
The spatial organization of sliding contacts also changes: during the first half of the simulation, sliding contacts are uniformly distributed throughout the packing, but in the second half, they become concentrated in certain regions.  This suggests that the loss of homogeneity occurs well before the appearance of shear bands.
During the second half events appear where the number of sliding contacts drops suddenly, and then rapidly recovers.
We show that these events are in fact local instabilities in the packing.
These events become more frequent as failure is approached.
For these two reasons,
we call these events precursors, since they are similar to the precursors recently observed in both
numerical \cite{Staron02,Nerone03} and experimental~\cite{Gibiat09,Scheller06,Zaitsev08,Aguirre06} studies of avalanches.

\end{abstract}

\section{Introduction}
\label{sec:intro}

The principal goal of many numerical studies of quasi-static granular materials
is to establish a connection between the macroscopic, continuum-like
response of the material, and its microscopic, grain-level state \cite{Luding04,Luding05,Kaneko03}.
One hopes to explain the macroscopic behavior using a small number of microscopic parameters that could then be used in a constitutive model.
The microscopic physical meaning of these parameters would provide the model with a physical justification, making it more general.
Several candidates have been examined,
such as the fabric tensor \cite{Madadi04}, force chains \cite{Cates98,Makse99} or force distributions \cite{Coppersmith96},
and sliding contacts \cite{Alonso05,Staron05},
In the limit of isostatic networks, stress paths have been calculated directly from the fabric tensor \cite{Gerritsen08}.
However, for arbitrary, frictional granular systems the subspace of all allowed force networks (and therefore stress paths) have to be included in the analysis \cite{Snoeijer06}. This leads to ambiguous values of the microscopic parameters.
Unfortunately, the future evolution of the system depends on these parameters.

Although the goal of a general, physically-based stress-strain relation
has not been attained,
several advances have been made in this direction.
One important point is the role of sliding contacts.
The contacts between the individual grains are assumed to be governed
by Coulomb friction.  
Thus each contact can be in one of two states: either ``sliding'' or ``non-sliding''.
The Coulomb friction law is sufficient to explain
the incremental non-linearity of granular materials:
previous studies \cite{Garcia-Rojo05b,Alonso04} have shown that
increasing the load leads to an augmentation of the number of sliding contacts,
which in turn causes the stiffness of the material to decrease.
If the loading direction is reversed,
the majority of the sliding contacts become non-sliding,
leading to an abrupt stiffening.
For this reason,
it is believed that the contact status (sliding or non-sliding) is the 
most relevant microscopic variable.

This connection between softening and sliding contacts
leads to the expectation that the number of sliding contacts
will continue to increase, right up to the time when the packing fails.
Although this expectation seems reasonable, it has not been checked.
Most numerical studies have focussed on cyclic loading
far from failure
\cite{Alonso04,Pena05},
or the critical state \cite{Zhang05,Unger05}.
In this paper, we examine granular packings under increasing load
up to the time of failure.
We find that the expectation of a monotonically increasing
number of sliding contacts is false.
The number of sliding contacts attains a clear maximum well before failure.
Thus the density of sliding contacts cannot be used as an internal variable.
Packings with the same number of sliding contacts may be in very different states.

Our work also shows the existence of precursors in biaxial tests.
Previously, precursors of avalanches have been identified
in both numerical \cite{Staron02,Nerone03} and experimental \cite{Zaitsev08,Scheller06,Gibiat09,Aguirre06} studies.
In these studies, the inclination of a static granular bed is slowly increased,
until an avalanche occurs.
Preceding the avalanche are numerous local reorganizations of the packing.
These events become more and more frequent as the angle of inclination is increased.

In biaxial tests,
precursors also become more frequent when the failure is approached.
They are triggered by changes in a localized region, and
they produce sound waves that propagate outwards.
At the origin of a precursor there is always a local instability that
triggers the precursor.
These instabilities resemble those that trigger failure in very small packings \cite{Welker09}.
The role of the precursors in the failure of large assemblies has yet to be investigated.

After a brief description of the simulation method,
we introduce to the simulation parameters 
and define the quasi-static limit in Sec.~\ref{sec:numproc}.
Next we give a description of the simulation in Sec.~\ref{sec:description}. We show how the stress-strain curve (Sec.~\ref{sec:stress_strain})
and the kinetic energy (Sec.~\ref{sec:KEvibrations}) evolve as the system approaches the failure.
Furthermore we examine the volume, the injected power, and the number of contacts in Sec.~\ref{sec:MVol}.
Thereafter we discuss in Sec.~\ref{sec:Csliding}
the evolution of the number of sliding contacts (Sec.~\ref{sec:Ms}),
the average force transmitted at sliding contacts (Sec.~\ref{sec:Cforce}),
the contact status transitions (Sec.~\ref{sec:ms_understanding},
and the spatial organization of sliding contacts (Sec.~\ref{sec:ttest}).
Last but not least, we discuss the two regimes of qualitatively different granular behavior prior to the failure (Sec.\ref{sec:tworegions}).
In a further section (Sec.~\ref{sec:precursor}),
we examine a precursor carefully, analyzing
the number of sliding contacts (Secs~\ref{sec:prec:Ms},~\ref{sec:prec:msdecr}), the stress-strain relation (Sec.~\ref{sec:prec:exp}), the stability of the packing (Sec.~\ref{sec:prec:instab}) as well as the evolution of the kinetic energy (Sec.~\ref{sec:precvel}). 
After that we show that the sliding contacts tend to cluster when a precursor appears, and the clusters disappear again afterwards (Sec.~\ref{sec:prec:msdecr}).
We show that the precursor is localized in the packing, but that the vibrations that appear afterwards travel through the packing.
The section is terminated by a summary of the precursor results in Sec.~\ref{sec:prec:summary}.
We conclude our work in Sec.~\ref{sec:conclusion} by some speculations on the significance of precursors for the failure.

\section{Numerical procedure}
\label{sec:numproc}

\subsection{Contact model}
\label{sec:Pmodel}
Grains are modeled as disks, and their interactions are calculated using the
common ``soft-sphere molecular dynamics'' method \cite{Cundall79}. The force at the grain
contact is generated by a linear dissipative spring whose length is given by the
overlap distance $D_n$:
\begin{equation}
F_n=-k_nD_n-\gamma_n\dot{D_n}.
\label{eq:Fn}
\end{equation}
Here, $k_n$ is the length independent spring stiffness and the damping
coefficient $\gamma_n$ controls the energy dissipation. The overlap distance $D_n$ is
calculated from the radii $r_i$ and $r_j$ of the touching particles and their
positions $\mathbf{x}_i$ and $\mathbf{x}_j$, 
\begin{equation}
D_n=|\mathbf{x}_i-\mathbf{x}_j|-r_i-r_j.
\label{eq:Dn}
\end{equation}
When the surfaces of the two touching disks move relative to each other a second
force $F_t$ arises, directed tangent to the particle surfaces. In analogy with
the normal force defined in Eq. (\ref{eq:Fn}) we have
\begin{equation}
F_t=-k_tD_t-\gamma_t\dot{D}_t.
\label{eq:Ft}
\end{equation}
In our case
$k_t=k_n$ and $\gamma_t=\gamma_n$.
Determining the change in the tangential spring length $\dot{D}_t$ involves both translational movement and rotation of the
two touching particles $i$ and $j$,
\begin{equation}
\dot{D}_t=-r_i\omega_i-r_j\omega_j
 +\frac{r_i+r_j}{r_i+r_j-D_n}(\mathbf{v}_i-\mathbf{v}_j)\cdot \mathbf{t}\,.
\label{eq:tspring}
\end{equation}
Here, $\omega_i$ and $\omega_j$ are the angular velocities of the touching
particles and $\mathbf{t}$ is a vector tangent to the particle surfaces at the
point of contact.
The factor in front of the last term
is needed to account for the overlap of the particles.
\par
In our simulation we
allow only for repulsive forces, $F_n > 0$. 
We also enforce
the Coulomb condition at each contact.
\begin{equation}
\mu F_n\geq |F_t|.
\label{eq:coulomb}
\end{equation} 
At a given moment, contacts where the strict inequality holds are nonsliding
contacts, whereas contacts with $\mu F_n=|F_t|$ are sliding.
When two particles separate,
the contact ``opens'', which can be considered as a third possible contact status.
In the following, we abbreviate the possible statuses of the contacts by $O$ (open),
$S$ (sliding) and $C$ (closed or non-sliding).

Contact status changes,
i.e., transitions between ``sliding'' and ``non-sliding'' play an important role
in this paper.
A contact undergoes a transition $C \to S$ (non-sliding to sliding)
when applying Eq.~(\ref{eq:tspring}) would lead to a violation of Eq.~(\ref{eq:coulomb}).
The reverse transition ($S \to C$) occurs when $\dot D_t$ in Eq.~(\ref{eq:tspring}) changes sign
or $F_n$ increases.
The transitions $S \to O$ and $C \to O$ occur when two touching grains separate,
and the reverse transitions $O \to S$ or $O \to C$ occur when two grains come into contact.

\subsection{Units and parameters}
\label{sec:units} 
Three parameters are set to unity for all simulations: the particle density
$\hat\rho$, the initial system length $\hat L$, and the pressure $\hat p$. This
defines our system of units. In two dimensions, the unit of force is 
$\hat f = \hat p  \hat L$,
and the unit of energy is $\hat E = \hat p \hat L^2$.
The unit of mass is $\hat m = \hat \rho \hat L^2$, whereas the time is measured in units of $\hat t = \hat L\sqrt{\hat\rho/\hat p}$.
The spring stiffness has a value of
$k_{n,t}=1600 \hat p$. 
This leads to overlaps that are a small fraction of the radius:
on average, we have $D_{n,t}/R \approx 0.3\%$. The small overlaps avoid the creation of many additional contacts.
The damping is $\gamma_{n,t} = 0.19 \hat m / \hat t$,
and the Coulomb friction coefficient is $\mu=0.25$.
No gravity is applied to the particles.
Unless otherwise mentioned, our systems have $N=16384$ particles with average particle mass $\bar{m}=2.8\,10^{-5}\hat \rho \hat L^2$. The total mass of the system is therefore of order unity.

\subsection{Boundary conditions}
\label{sec:bc}
\begin{figure}
\centering
\includegraphics[width=0.6\columnwidth]{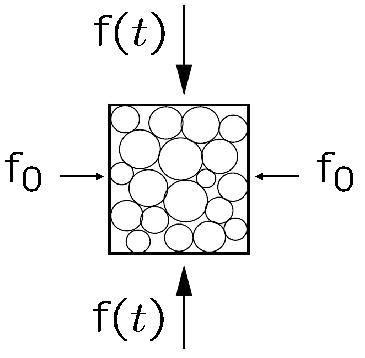}
\caption{Biaxial boundary conditions. The walls are smooth and can move normal to their surfaces.}
\label{fig:biaxial}
\end{figure}

We apply biaxial boundary conditions. These are easy to implement and simple to
handle. In each direction, the granular packing is delimited by a light-weight
($m=0.01\hat m$) moveable wall parallel to one of the coordinate axes, as
shown in Fig.~\ref{fig:biaxial}.  A force is applied to each boundary that can
be constant or time-dependent.  In this way, one can fix the average stress
inside the granular packing.  The deformation of the packing can be determined
by monitoring the movements of the boundaries.

In our setup,
the walls are perfectly slippery, i.e.
they exert only normal forces.
This has several advantages.
First, the forces on opposite walls are guaranteed to be equal.
Second, the number of degrees of freedom is reduced.
And last but not least, larger systems are more homogeneous.

\subsection{Preparation of initial conditions}
\label{sec:preparation}

The 2 dimensional granular medium is made up of disks with radii $r$ uniformly
distributed within the range $[0.7r_\mathrm{max}, 1r_\mathrm{max}]$ with $r_\mathrm{max}=3.5\,10^{-3}$ and initial velocity in the
range $v_x,v_y\in [-0.5, 0.5]$.
The initial radii and velocities are chosen using a random number generator,
and a series of packings is generated
by changing the ``seed''.
The resulting assemblies obtained essentially
differ in contact topology and the spatial distribution of the grain sizes.
In this paper, we study in detail the behavior of one packing containing $16384$ disks. However, we compared the behavior of this packing to three other packings of this size, and we found that the results are always the same. Therefore the results do not depend on the configuration but apply for all packings of this size.

The frictionless particles are initially separated,
but a constant external force $f_0=0.8 \hat f$ is
applied to each of the walls,
so that they move inward and compress the grains into a packing
of approximate size $0.8 \hat L$
capable of transmitting normal forces.  

During the compression process,
kinetic energy is removed by the damping at particle contacts.
Certain motions, however, require special care.
For example, the velocity of the center of mass
cannot be damped by contact forces,
and thus a global viscous damping is applied for $10\hat t$. 
The kinetic energy in the assembly decreases to $2.6\,10^{-11} \hat E$.
The main reservoir of remaining kinetic energy are particles
without contacts.
To remove their energy,
a viscous damping force opposing the individual grain movement is
then applied for $40\hat t$, and the remaining kinetic energy is
$4.6\,10^{-21} \hat E$.
At the end of the preparation, friction is turned on.

\subsection{Loading the sample}
\label{sec:loading}
The configurations obtained are submitted to an increasing external force
along the vertical axis, whereas the horizontally applied forces remain
unchanged.
The vertical force increase is linear,
\begin{equation}
f(t)=f_0+\alpha t
\label{eq:fext}
\end{equation}
To obtain a pressure $p = f_0/L$ that is approximately unity,
we choose $f_0=0.8\hat f$.
The prefactor $\alpha$ determines the value of the very small force increment
per timestep.
We study failure in the quasi-static limit,
meaning that the applied force at which the assembly fails
is independent of $\alpha$, in our case $\alpha=1.28\,\,10^{-2}\hat{f}/\hat{t}$.
We carefully checked that failure is independent of $\alpha$ for lower values of $\alpha$, but shifts to larger forces $f(t)$ for higher values.
At the beginning, a parabolic matching is applied to obtain a
continuous differentiable force curve.

\section{Description of the simulations}
\label{sec:description}

\subsection{Stress-strain curve}
\label{sec:stress_strain}

\begin{figure}
\centering
\includegraphics[angle=270, width=\columnwidth]{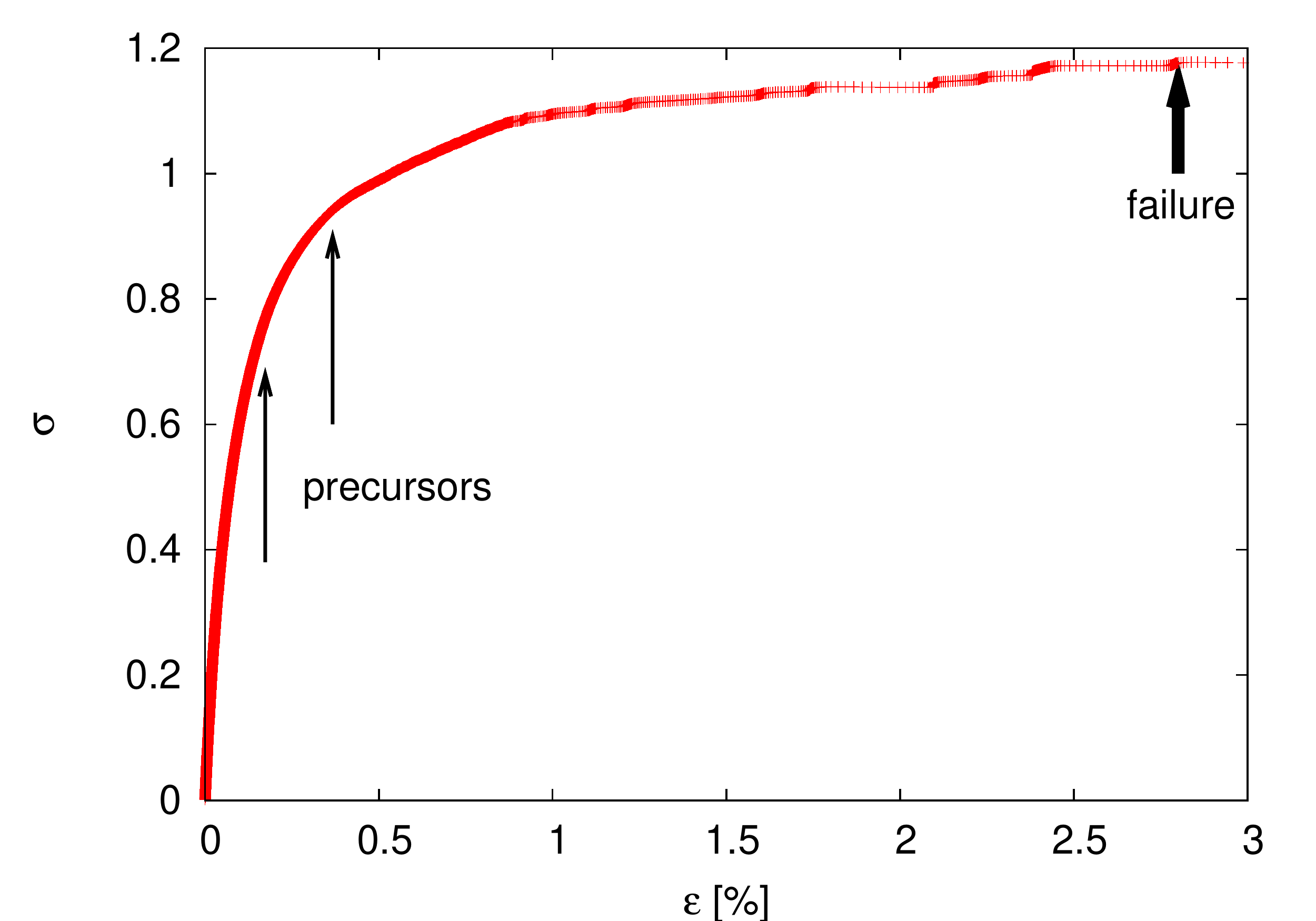}
\caption{Deviatoric stress versus deviatoric strain for a typical
quasi-static simulation with a large number of particles (parameters defined in Secs~\ref{sec:units},~\ref{sec:loading} and discussed in Secs~\ref{sec:loading},~\ref{sec:stress_strain}).
The two slim arrows indicate two precursors which are also indicated in Figs~\ref{fig:KE},~\ref{fig:sliding},~\ref{fig:FN} and~\ref{fig:ms_changes}. The heavy arrow indicates the beginning of the failure. When failure happens, the strain jumps to a value close to 22\%.}
\label{fig:stress_strain}
\end{figure}

Fig.~\ref{fig:stress_strain} shows a typical stress-strain curve for one simulation.
Here, the deviatoric strain is defined as
\begin{equation}
\epsilon = \frac{L_x - L_{x0}}{L_{x0}} - \frac{L_y - L_{y0}}{L_{y0}}\,,
\end{equation}
Where $L_x,L_y$ are the horizontal and vertical length of the system in Fig.~\ref{fig:biaxial}, and $L_{x0},L_{y0}$ are their initial values. Similarly, the deviatoric strain is
\begin{equation}
\sigma = \frac{f(t)}{L_x} - \frac{f_0}{L_y}\,.
\end{equation}
We will refer to the slope of the curve in Fig.~\ref{fig:stress_strain} as the "stiffness of the assembly".  At the beginning, the assembly is very stiff (the stiffness is comparable to $k_n=1600$, which is the stiffness of one contact).
As the stress increases, the assembly becomes softer.
The thin arrows indicate two ``precursors''  or small rearrangement events preceding failure. These events leave no clear sign on the stress-strain graph, but appear clearly when other quantities are plotted. We will examine the precursors in detail in Sec.~\ref{sec:precursor}.
Finally, the slope of the curve becomes nearly flat, and the assembly is very weak. Then little horizontal jumps appear that are 
associated with rearrangement processes.
These are events where the walls move rapidly. The heavy arrow indicates the beginning of failure, which we define as the event where the strain crosses a threshold of 5\%. This value is more than twice the strain just before failure, but much lower than the strain after failure.

We performed several tests on systems with different numbers $N$ of particles, but all having the same mass and size.  We find that the mean value of $f_\mathrm{fail}$ varies only slightly with $N$, whereas its variance decreases significantly for larger systems.
This indicates that $f_\mathrm{fail}$ is independent of system size.

\subsection{Kinetic energy and vibrations}
\label{sec:KEvibrations}

\begin{figure}
\centering
\includegraphics[angle=270, width=\columnwidth]{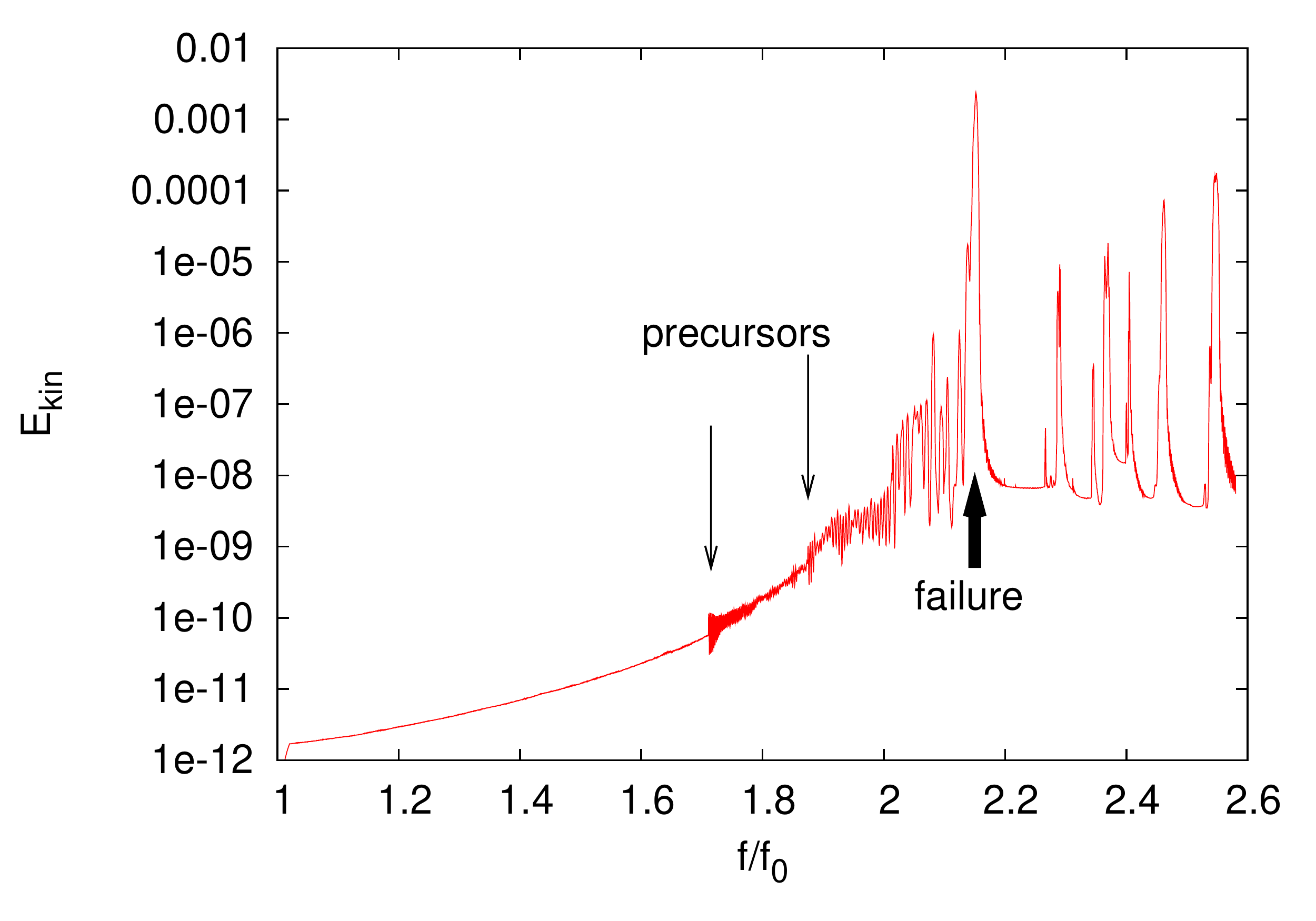}
\caption{Kinetic energy for one simulation at $\alpha=1.28 \,\, 10^{-2} \hat E/\hat t$. At the failure, $E_\mathrm{kin}$ rises by many orders of magnitude (heavy arrow).}
\label{fig:KE}
\end{figure}

During the simulations, the kinetic energy $E_\mathrm{kin}$ rises as the packing becomes softer due to some contacts becoming sliding and some contacts disappearing.
This behavior is shown in Fig.~\ref{fig:KE}. For instance, a two-decade increase of $E_\mathrm{kin}$ is accompanied by a
one-decade decrease of stiffness.
When the packing becomes very soft, vibrations become much larger in amplitude.
The two precursors of from Fig.~\ref{fig:stress_strain} are indicated by arrows but not visible in the figure, for their duration is comparable to the distance between two consecutive data points.
On the other hand they provoke slowly damped vibrations visible in Fig.~\ref{fig:KE}.
The typical oscillation period is $T=2\,10^{-3}\hat t$.
Finally failure shows up as a pronounced maximum of kinetic energy at about $t=74\hat t$ ($\sim 4\,10^{4}$ oscillation periods).

\subsection{Change of volume, power injected and the number of contacts}
\label{sec:MVol}
\begin{figure}
\centering
\includegraphics[angle=270, width=\columnwidth]{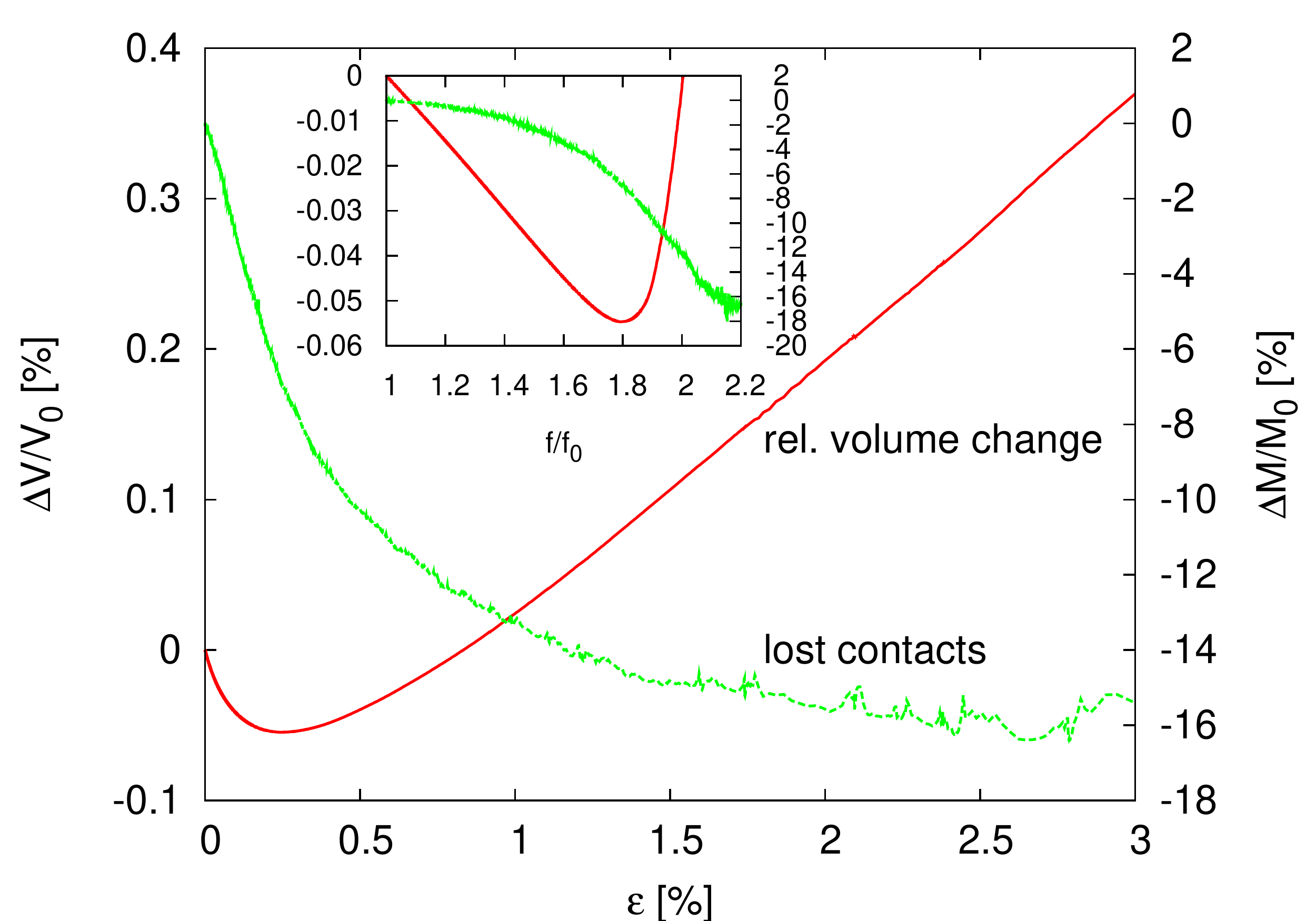}
\caption{Relative change of the volume (left scale) and relative change of the number of contacts (right scale) as a function of deviatoric strain. The inlay shows the same quantities as a function of external force.}
\label{fig:MVol}
\end{figure}

One expects the volume of the packing to decrease with the increase of $f_\mathrm{ext}$ (Fig.~\ref{fig:MVol}). This decrease is very small for the stiff particles system ($\Delta V_\mathrm{max}/V_0=-0.05\%$ at $\epsilon=0.25\%$). With the decrease of $V$, one expects the number of contacts $M$ to rise. However, we observe a \textsl{decrease} in $M$ during the same strain interval.
The decrease of $M$ means that the particles try somehow to avoid each other.
It should be kept in mind that
the systems we prepare are close to densest packing, as no friction is applied during preparation.
Turning on friction after preparation might provide the reason for the decrease of $M$, as some particle motions will be blocked.
After $\epsilon=0.25\%$ the volume starts to increase again, while $M$ still decreases, but at a lower rate. Note that the increase in $V$ is almost linear in $\epsilon$. The rate of the decrease in $M$ slows further down with $\epsilon$ until vanishing at the failure.

\begin{figure}
\centering
\includegraphics[angle=270, width=\columnwidth]{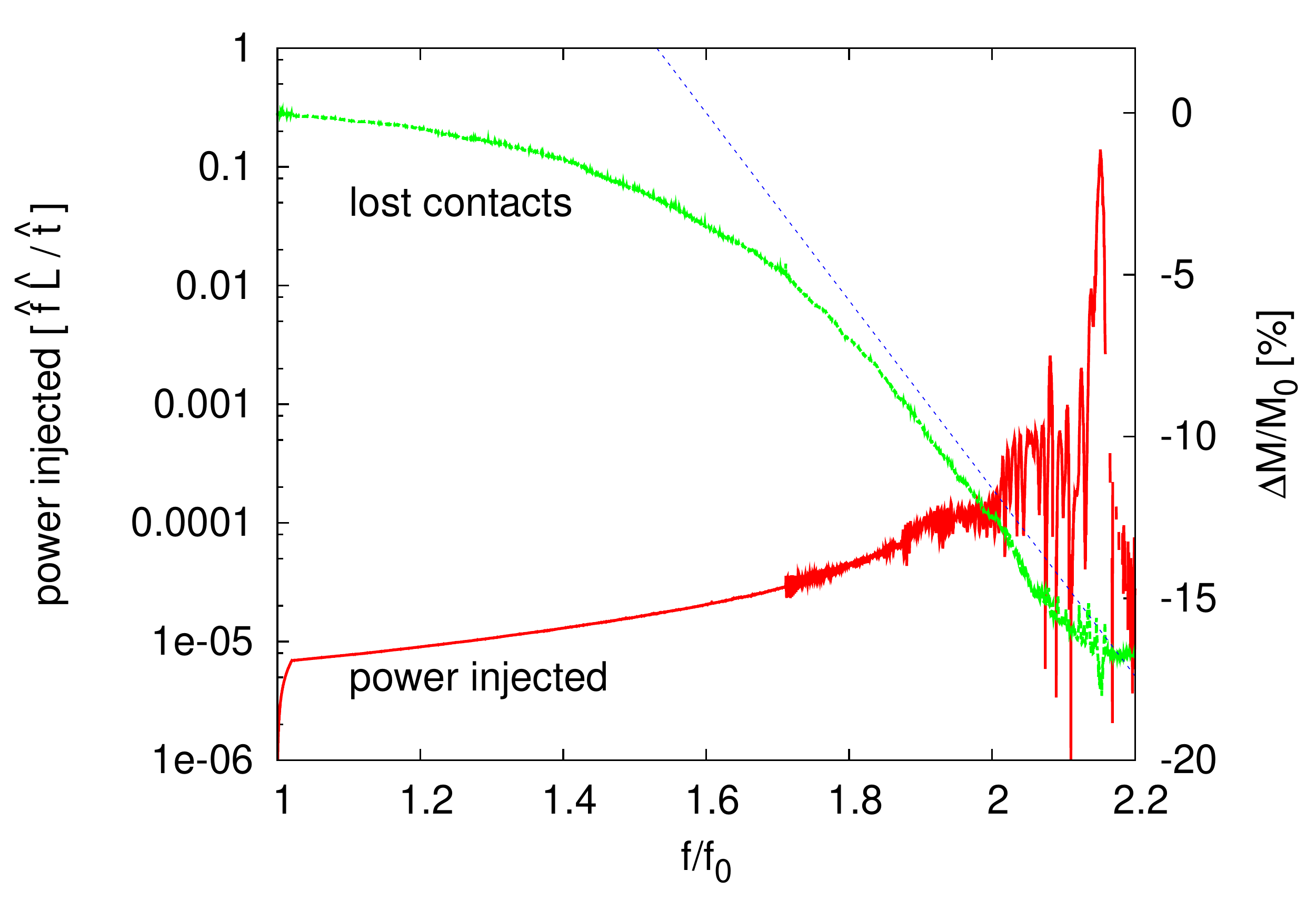}
\caption{Power injected into the system (left scale) and number of contacts (right scale) as a function of external force. The power injected per unit of time increases quickly after $f/f_0=1.8$.
Simultaneously, the number of contacts decreases linearly (the dashed line is a guide to the eye).}
\label{fig:power}
\end{figure}

Therefore we have two regimes in Fig.~\ref{fig:MVol}: a decrease in volume until $\epsilon=0.25\%$, and thereafter an increase in $V$ until the failure. An increase in volume usually means that the system is loosing energy, as power has to be injected to decrease the volume. However, when we calculate the power injected at the walls (Fig.~\ref{fig:power}) we see that is is always positive: the energy of the system always rises.
Note that this behavior is unexpected: the system should simply explode.
The key to resolving this paradox is that the stress is strongly anisotropic: $f\sim 2f_0$.
We will show that there are many qualitative differences in the systems behavior between the two regions in Sec.~\ref{sec:tworegions}.

A loss in contacts $M$ also always implies a loss of additional possibilities to stabilize the packing: when fewer contacts are present, it is more difficult for the system to balance a higher load. Figure~\ref{fig:stress_strain} shows that the higher the actual load, the more the system deforms with a further increase.
This evidences that a lower number of contacts correlates with a higher deformation.

Besides the number of contacts, sliding contacts require special care. These are detailed in the next section.

\subsection{Sliding contacts}
\label{sec:Csliding}

\subsubsection{Number of sliding contacts}
\label{sec:Ms}
\begin{figure}
\centering
\includegraphics[angle=270, width=\columnwidth]{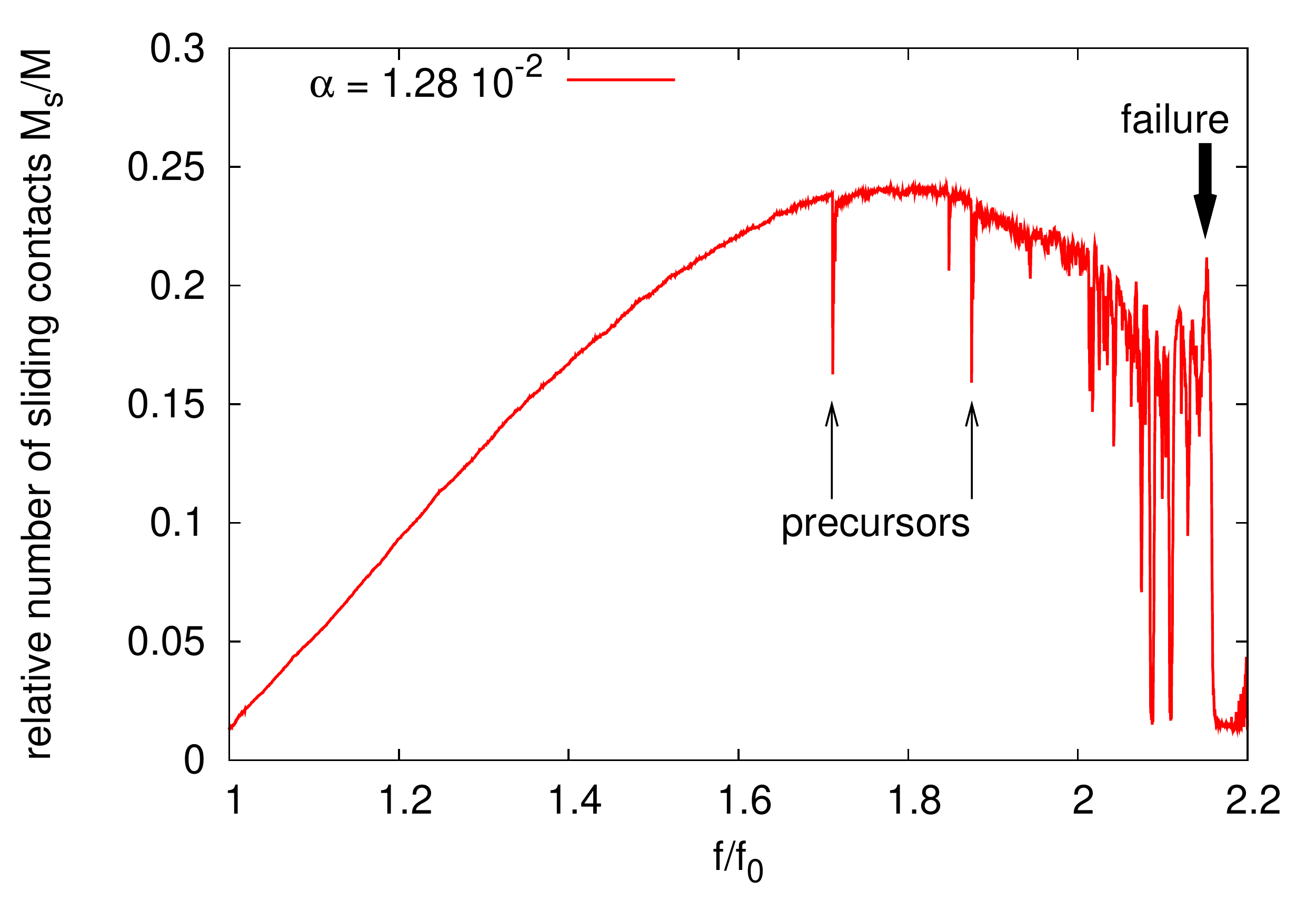}
\caption{
The number of sliding contacts $M_s$.
There is always a decrease of $M_s$ before failure, followed by an increase at the failure. Right after the failure $M_s$ is close to zero as only the contacts with the walls remain sliding ($\mu_\mathrm{wall}=0$).
}
\label{fig:sliding}
\end{figure}
The number of sliding contacts rises as the deviatoric part of the external stress on the packing is increased (Fig.~\ref{fig:sliding}).
Previous works
\cite{Garcia-Rojo05b,Alonso04,Welker09,Garcia-Rojo05} 
lead one to suppose that the number of sliding contacts finally attain a maximum at failure.
However, this is not true, as
the maximum is attained well before failure (Fig.~\ref{fig:sliding}).
This maximum coincides with the minimum in volume, shown in the inlay of Fig.\ref{fig:MVol}.
In small packings the existence of a maximum before the failure is not observed, probably because of the limited number of contacts. This small number allows only for few contact status changes at a time, and every change results in large changes in the stability of the granular assembly. Therefore very few changes can already lead to instability and cause failure \cite{Welker09}.

We observe that when the increase in $M_s$ becomes slower, the number of sliding contacts plunges at certain positions and then quickly recovers.
The sudden plunges in the number of sliding contacts are the signature of precursors.  In much faster simulations ($\alpha\gg 10^{-2}$ in Eq.~\ref{eq:fext}), they are not observable.
We carefully checked this on the basis of an equal number of data points for many simulation speeds.

During failure,
large rearrangements occur, and almost all
sliding contacts between particles disappear or close.

\subsubsection{Strength of Sliding Contacts}
\label{sec:Cforce}
\begin{figure}
\centering
\includegraphics[angle=270, width=\columnwidth]{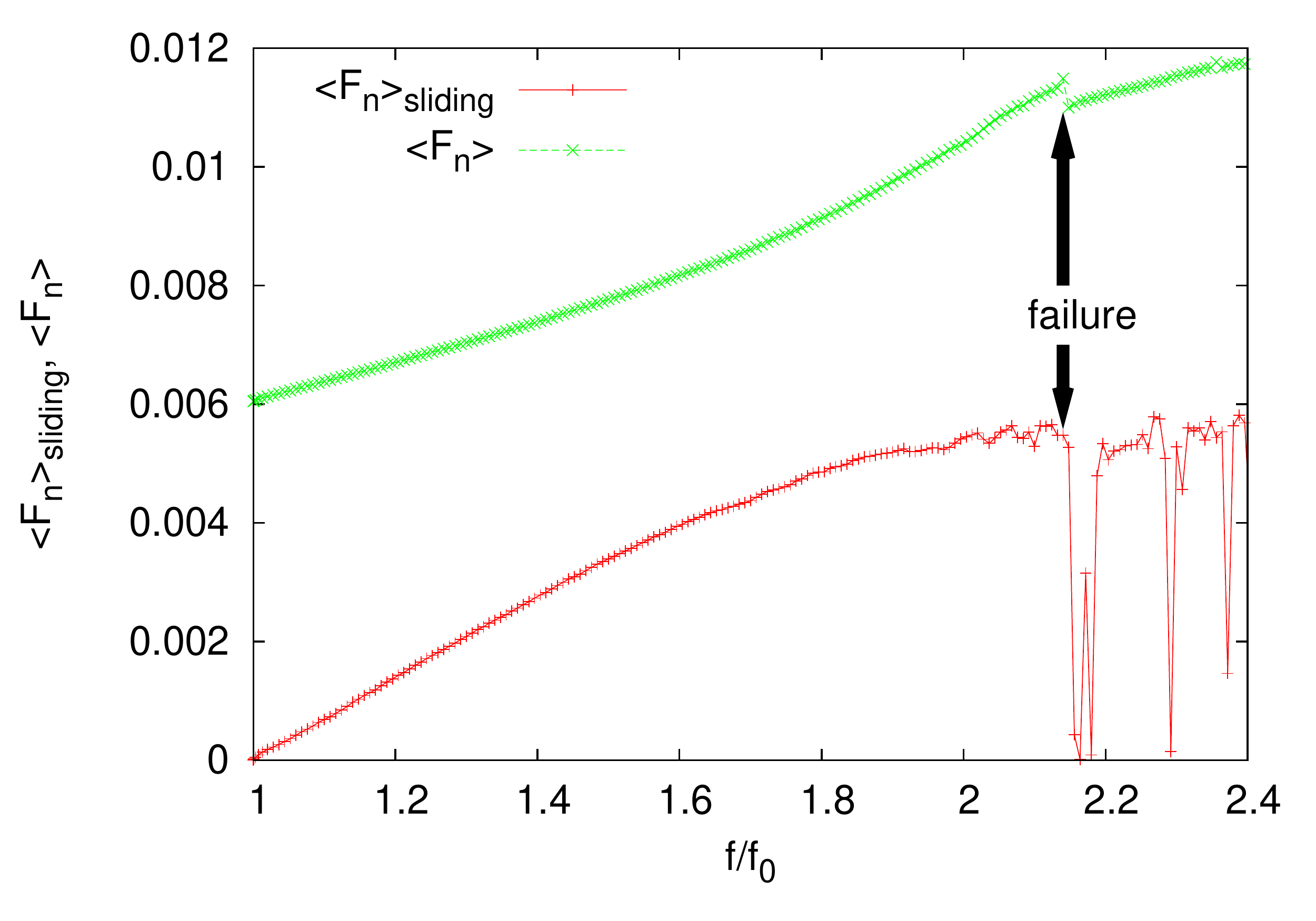}
\caption{Average normal force transmitted at a contact, and average normal force transmitted at sliding contacts}
\label{fig:FN}
\end{figure}

Fig.~\ref{fig:FN} shows the average normal force $\langle F_n\rangle$ at a contact and compares its value to the average normal force at a sliding contact $\langle F_n\rangle _\mathrm{sliding}$.
$\langle F_n\rangle _\mathrm{sliding}$ is always much smaller than $\langle F_n\rangle$,
in accord with the findings of Ref.~\cite{Radjai98}: most sliding contacts transmit less than average forces.
Both $\langle F_n\rangle$ and $\langle F_n\rangle _\mathrm{sliding}$ increase about linearly in the first half of the simulation. However, the average force at sliding contacts increases much faster than the average force for all contacts.
Then, later in the simulation, $\langle F_n\rangle$ increases faster than before, whereas the increase in $\langle F_n\rangle _\mathrm{sliding}$ slows until a plateau is reached close to the failure.

Note that in the beginning of the simulation
the average force at sliding contacts is very small. Therefore the first appearing sliding contacts are very weak contacts. As the deviatoric stress is increased, new sliding contacts appear that carry larger forces.
When the maximum in $M_s$ is reached, the increase in $\langle F_N\rangle _\mathrm{sliding}$ becomes slower. Finally $\langle F_n\rangle _\mathrm{sliding}$ attains a maximum when $M_s$ starts to decrease again.
This leads to the conclusion that while sliding contacts constantly disappear, this does not change the average force at sliding contacts. This finding implies that sliding contacts are not part of the force chains carrying the external load.
On the other hand the global mean $\langle F_n\rangle$ increases strongly close to the failure. We will see in the next subsection that this behavior is linked to the disappearance of contacts, so every remaining contact has to carry a higher fraction of the load, on average.

\subsubsection{Understanding the evolution of $M_s$}
\label{sec:ms_understanding}
We will now examine in detail the variation in $M_s$. We will see that both the increase and the following decrease can be understood
in terms of the frequency of transitions between the three contact statuses that have been introduced in Sec.~\ref{sec:Pmodel}: closed, sliding, and open. 

\begin{figure}
\centering
\includegraphics[angle=270, width=\columnwidth]{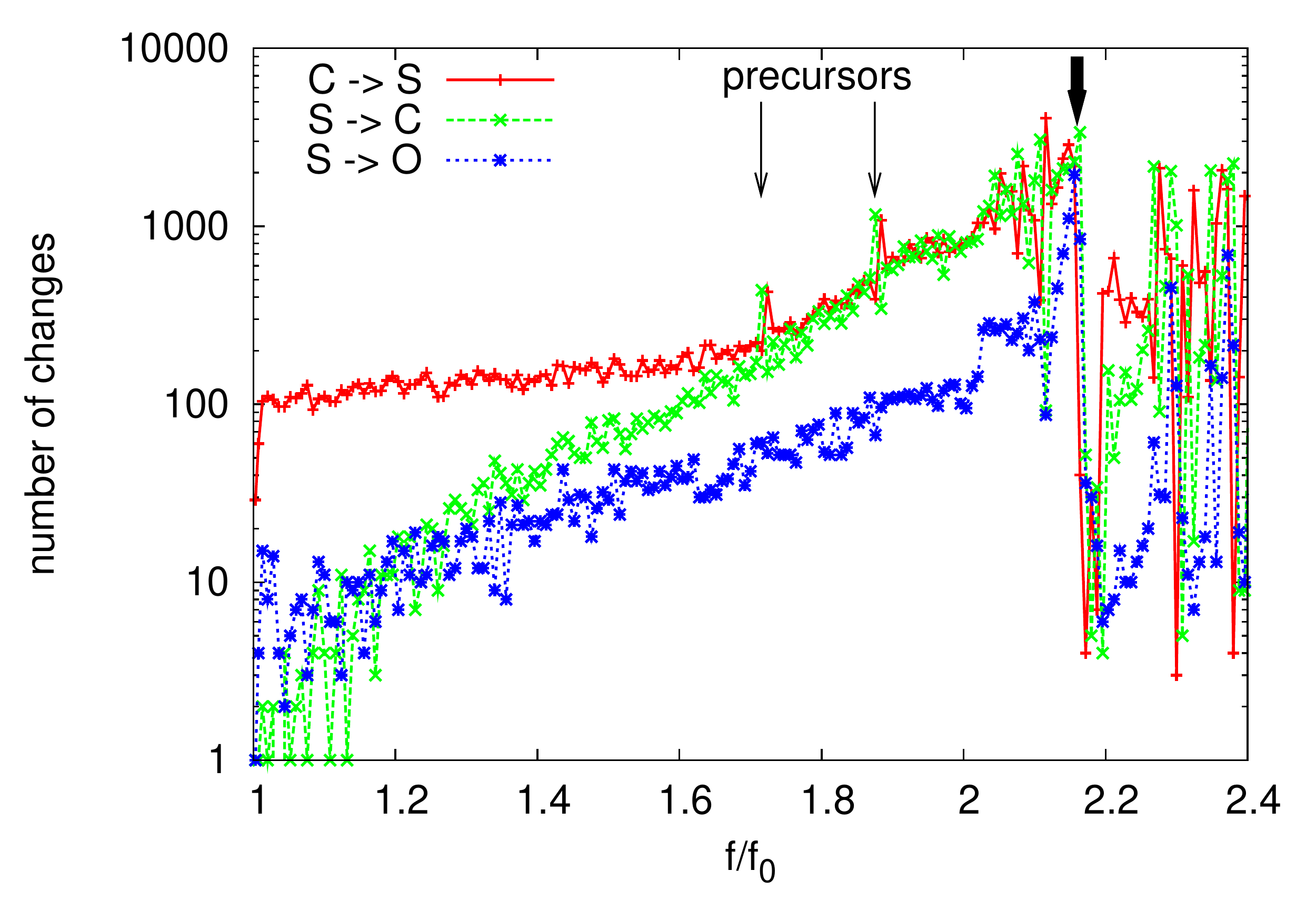}
\caption{Contact status changes involving sliding contacts ('S'). The heavy arrow indicates the failure. See Sec.~\ref{sec:ms_understanding} for details.}
\label{fig:ms_changes}
\end{figure}

In Fig.~\ref{fig:ms_changes} we show the most important contact status changes.
In the beginning the most frequent transitions are from closed to sliding 'C$\rightarrow$S'. But the number of inverse transitions 'S$\rightarrow$C' increases exponentially,
and becomes approximately equal to the frequency 'C$\rightarrow$S'
at about $f/f_0=1.7$, i.e., near the maximum of $M_s$.
From then on, the two transitions cancel each other, i.e.
their difference fluctuates around zero.
The third most important transition is 'S$\rightarrow$O'. Once 'S$\rightarrow$C' and 'C$\rightarrow$S' cancel each other, 'S$\rightarrow$O'
leads to the decrease in $M_s$ that is observed in Fig.~\ref{fig:sliding} before the failure.
The other possible contact status changes (not shown in the figure) are much less frequent. At the failure itself at $f/f_0\approx 2.1$, all sorts of contact changes become very important, even those that are not displayed.
The signature of two precursors can also be seen in the figure:
sharp peaks
in 'C$\rightarrow$S' and 'S$\rightarrow$C' at $f/f_0\approx 1.75$ and $1.9$ (See the next section for details on precursors).

Another important question is the spatial organization of sliding contacts, which we will investigate in the following.

\subsubsection{Spatial distribution of sliding contacts -- ordering effects close to the failure}
\label{sec:ttest}

To assess the spatial organization of sliding contacts,
we investigate whether a Poisson process \cite{Getis79}
could generate their observed spatial distribution.
Recall that a Poisson process is one where a
fixed
number of points
in a region
are selected,
with each point having an equal probability of being chosen.
Furthermore, each point is chosen independently of the others:
the choice of point $A$ has no influence on the probability of choosing point $B$.
If the
region is subdivided into boxes of equal size,
the probability of observing $x$ points in a box is
\begin{equation}
P(x;\lambda) = \frac{e^{-\lambda}\lambda^x}{x!},\quad x=0,1,2,\ldots
\label{eq:Poisson}
\end{equation}
Here, $\lambda$ is the average number of points expected in one box.
Note that the variance of the Poisson distribution is equal to its mean.
We will make use of this later.

To check if the sliding contacts are distributed according to a Poisson process,
we divide the packing into equally-sized square boxes of length $l=4d$ ($d$ is the average particle diameter), and count the number of sliding contacts in
each box. One can then compare the observed frequencies with the prediction in
Eq.~(\ref{eq:Poisson}).  One convenient way to do this is the so-called ``$t$-test''.

\begin{figure}
\centering
\includegraphics[angle=270, width=\columnwidth]{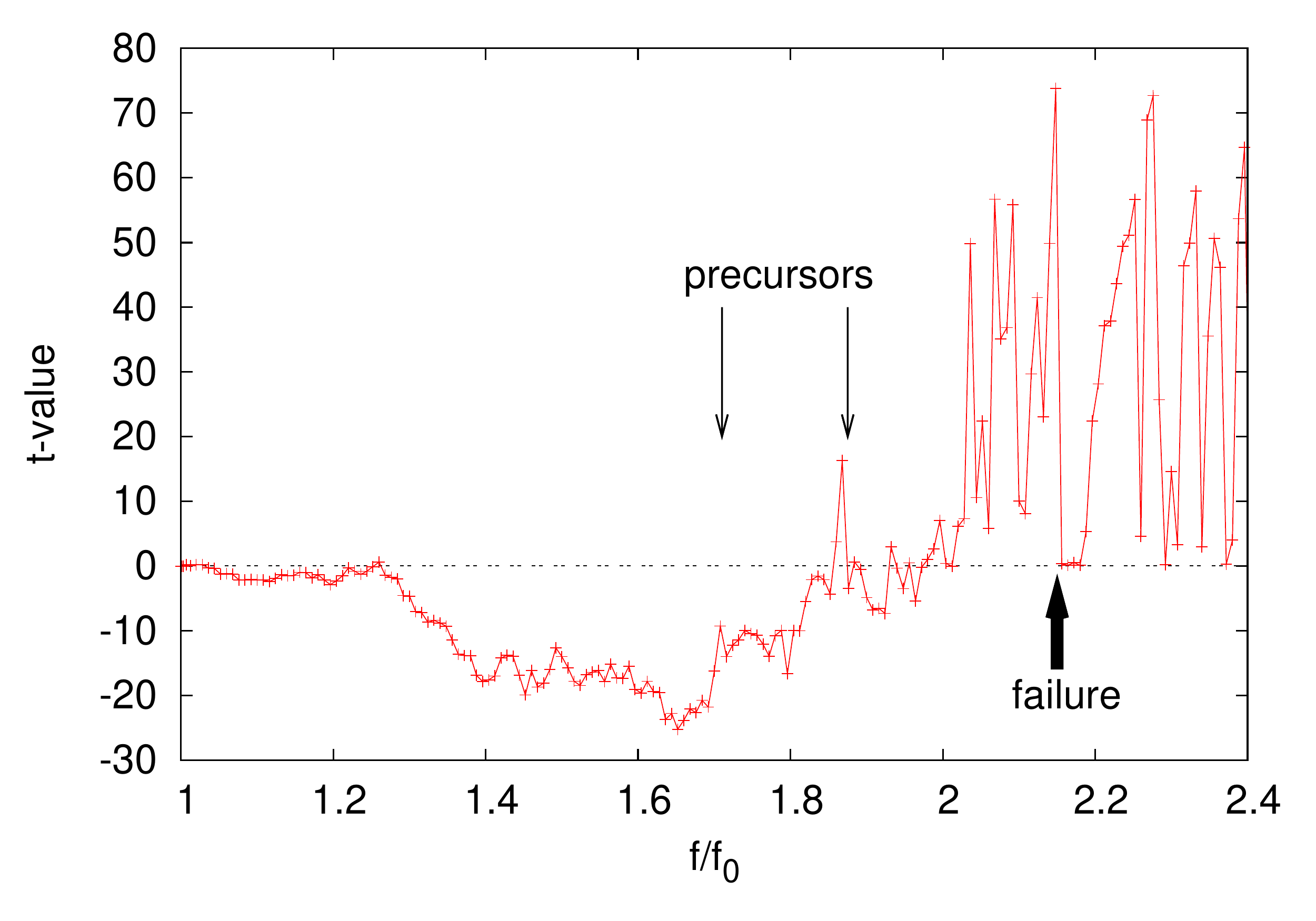}
\caption{Values from the $t$-test (Eq.~\ref{eq:ttest}).
This test checks the spatial distribution of sliding contacts:
$t<0:$ contacts are more uniform than random,
$t>0:$ contacts cluster.
}
\label{fig:ttest}
\end{figure}

The t-test compares the variance $\sigma^2$ of the distribution to the mean $\bar{M}_{s,box}$ \cite{Getis79}:
\begin{equation}
\displaystyle
t=\frac{(\sigma^2-\bar{M}_{s,box})}{\sqrt{2/(N-1)}}
\label{eq:ttest}
\end{equation}
Negative $t$-values indicate low variance or evenness of the distribution while positive $t$-values indicate a departure in the direction of high variance or clumping (clustering).
The values in Fig.~\ref{fig:ttest} show that in the beginning of the simulation the sliding contacts tend to be distributed randomly over the packing ($t\approx 0$). However, the very initial values might not be significative as the number of sliding contacts is small.
As more sliding contacts appear, the t-value decreases and becomes negative, indicating a sharply peaked distribution (low variance), meaning that the sliding contacts repel each other: the presence of one sliding contact reduces the probability that a neighboring contact will become sliding.
Later, close to the failure,
the $t$-values become positive and increase strongly. This indicates that at the failure the sliding contacts strongly tend to cluster.

\begin{figure}
\centering
\includegraphics[angle=270, width=\columnwidth]{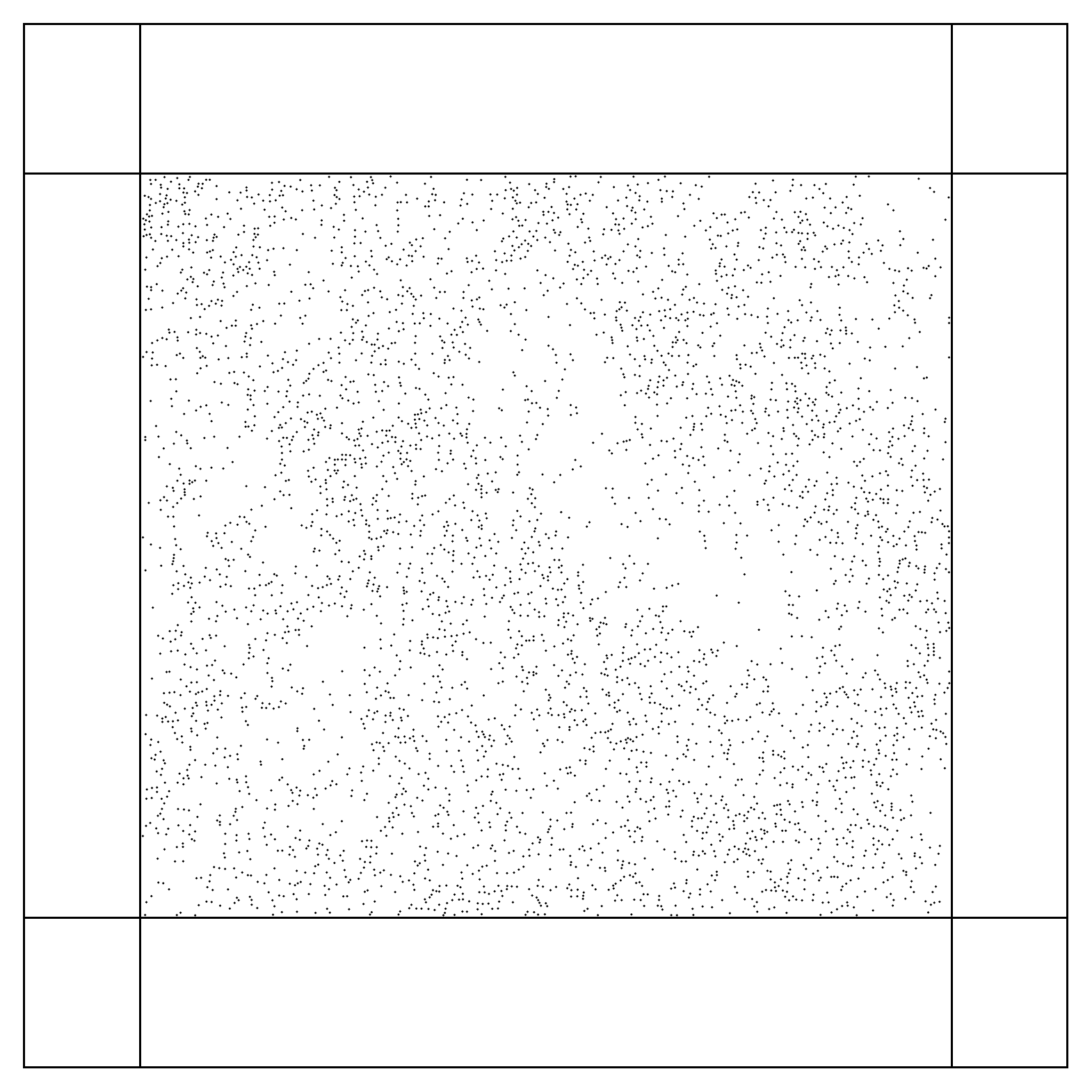}
\caption{Positions of the sliding contacts for $f/f_0=2.15$, i.e., just before failure.
At this time, the $t$-test yields $t = 50$.
}
\label{fig:Ms_in_packing}
\end{figure}
Fig.~\ref{fig:Ms_in_packing} shows the positions of the sliding contacts just before failure (at $f/f_0 = 2.15$, failure is at $f/f_0 = 2.18$).  One discerns a diffuse diagonal band that crosses the sample from lower left to upper right.  At failure, a shear band forms in this region.
The other sliding contacts are concentrated at some distance in another band that is parallel to the shear band. The formation of the shear band will finally lead to the failure of the assembly.

\subsection{Two qualitatively different regimes}
\label{sec:tworegions}

As discussed in Sec.~\ref{sec:MVol}, we can identify two regimes of loading with different behavior. Table~\ref{tab:tworegions} shows the behavior during the two periods. Most quantities in the table depend on the contact status changes. In the first period ($f/f_0< 1.8$) the most frequent contact transition is from closed to sliding. Therefore the number of sliding contacts $M_s$ increases with the load, while their spatial distribution is quite uniform ($t<0$). Also the volume decreases during this period. In the second regime, the dominant contact status transition is from sliding to open (disappearing of formerly sliding contacts). This entails a decrease in $M_s$, and a clustering tendency ($t>0$). During this period, the average normal force transmitted at sliding contacts does not increase any more. Last but not least, the change in volume is reversed: $V$ increases, while the power injected stays positive (see Fig.~\ref{fig:power} in Sec.~\ref{sec:MVol}).

Towards the beginning of the second regime, precursors of failure appear that become more frequent with increasing $f$.
Their appearance has been outlined in Sec.~\ref{sec:Ms} and indicated in Fig.~\ref{fig:sliding}. They will be discussed in detail in the next section.
\begin{table}
\centering
\begin{tabular}{c||cc}
		& $f/f_0<1.8$	& $f/f_0>1.8$\\\hline\hline
dominant status change	& C$\rightarrow$S	&	S$\rightarrow$O\\
$M_s$	&	$\nearrow$		&	$\searrow$\\
$t$-test ($M_s$)			&	$<0$				& $>0$				\\
$\langle F_n \rangle _\mathrm{sliding}$	& $\nearrow$ &	$=$\\
V		& $\searrow$		& $\nearrow$ \\
\end{tabular}
\caption{Time behavior of characteristic quantities during the two periods. The arrows indicate the evolution with increasing load (rising $\nearrow$ or falling $\searrow$).}
\label{tab:tworegions}
\end{table}

\section{Precursors}
\label{sec:precursor}

\subsection{Definition}
\label{sec:prec:def}
Prior to the collapse of the packing, several precursors occur where $M_s$ plunges and then quickly recovers.
We define the precursor to be an event where $M_s$ plunges by at least 10\% of its maximum value before the failure.
This drop in M$_\mathrm{s}$ varies from precursor to precursor, but the qualitative behavior of the precursors is always the same.
A closer inspection in this section will show that precursors are initiated by an instability that gives rise to a local increase in the kinetic energy. The decrease in the number of sliding contacts is then just a consequence of the release of potential energy.

Figure \ref{fig:sliding} tells us that the precursors become more frequent as the failure is approached.
Therefore they might play an important role for the appearance of failure.
In the next section we will see what happens at one precursor.

\subsection{Examination of a Precursor}
\label{sec:prec:examine}

\subsubsection{Number of sliding contacts}
\label{sec:prec:Ms}
To better understand the precursors, we examine in detail the first precursor indicated by the first arrow in Figs~\ref{fig:stress_strain},~\ref{fig:KE},~\ref{fig:sliding},~\ref{fig:ms_changes} and~\ref{fig:ttest} at $f/f_0 \approx 1.71$.
Figure~\ref{fig:KEpre} shows $M_s$ around this precursor.
At its appearance, the drop in $M_s$ is very sharp, while the recovery afterwards is slower and represents the relaxation to a new (force) equilibrium.
Figure~\ref{fig:KEpre} also shows the kinetic energy at that precursor. When $M_s$ starts to decrease, $E_\mathrm{kin}$ increases quickly. However, the kinetic energy very soon decreases again. This happens \textsl{before} $M_s$ reaches the minimum value. The maximum of $E_\mathrm{kin}$ can vary from one precursor to another one, but it is always much smaller than the maximum at failure. This is due to the limited time in which the energy rises \cite{Welker09}.

\subsubsection{Appearance of an instability}
\label{sec:prec:instab}
\begin{figure}
\centering
\includegraphics[angle=270, width=\columnwidth]{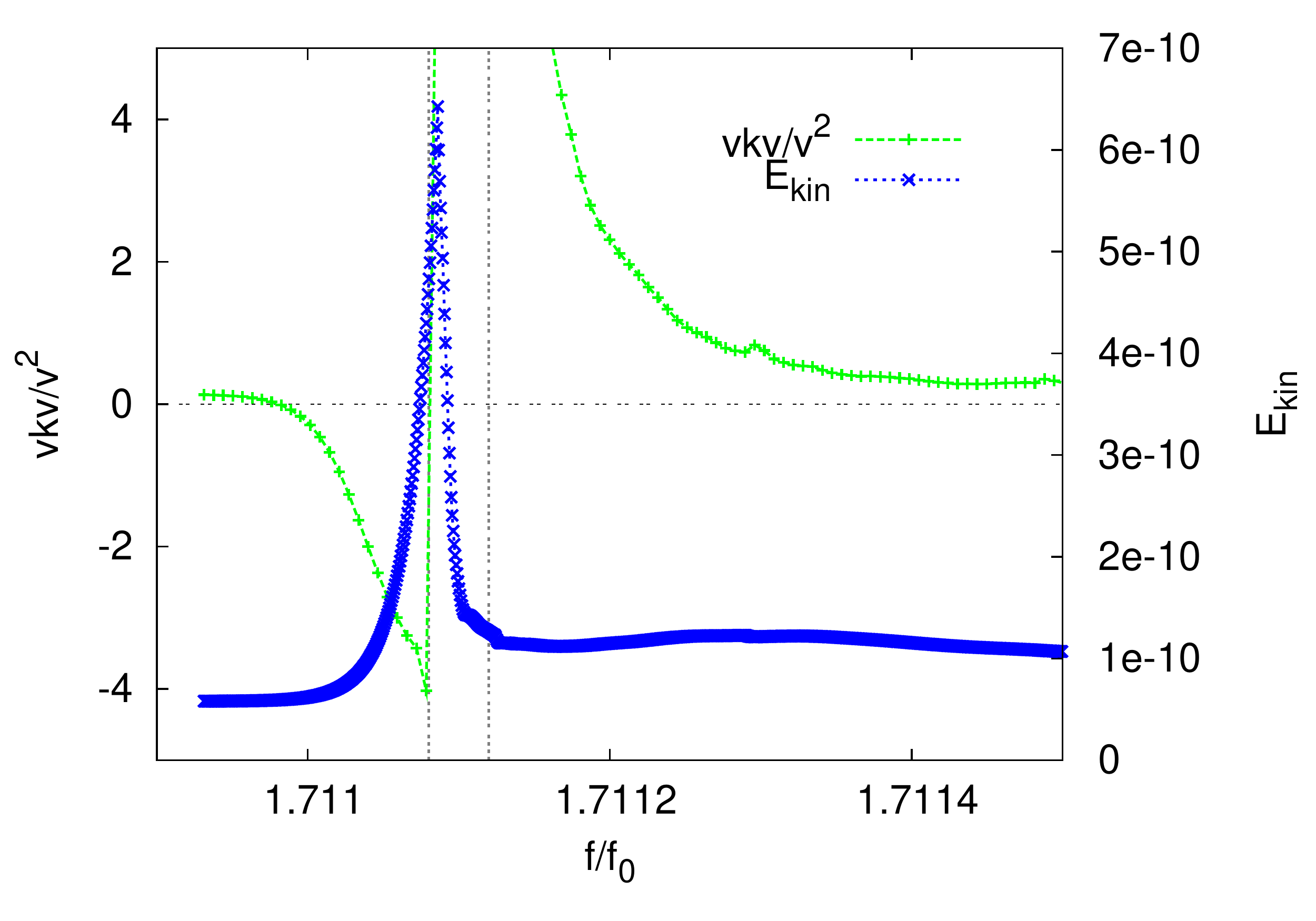}
\caption{Stability of the assembly at the occurrence of the precursor.
When the instability appears, $E_\mathrm{kin}$ rises exponentially.
The dotted horizontal line separates the two regions `stability' and `instability'. The two vertical lines indicate the positions of Figs.~\ref{fig:KE1},\,\ref{fig:KE2}.}
\label{fig:vkvpre}
\end{figure}

In Fig.~\ref{fig:vkvpre} we 
show another measure of the stiffness
of the assembly.
This stiffness is calculated by reducing the stiffness matrix $\mathbf{k}$ (see appendix \ref{app:vkv}), which contains the stiffnesses of all the contacts, to a scalar stiffness by multiplication with the particle velocities $\mathbf{v}$:
\begin{equation}
k=\mathbf{vkv}/\mathbf{vv}\,.
\end{equation}
Note that $k$ contains the velocities of all the particles, whereas the stiffness defined earlier in Sec.~\ref{sec:stress_strain} concerns only the walls.
The advantage of $k$ is that it can detect localized instabilities \cite{Welker09}.
Specifically, $k<0$ means that the packing is (at least locally) unstable, whereas $k>0$ indicates that it is stable.  Furthermore, $k$ is correlated to the stiffness defined in Sec.~\ref{sec:stress_strain}: large positive $k$ correspond to stiff assemblies.
In Fig.~\ref{fig:vkvpre} we see the stability of the assembly at the time when the precursor appears. At $f/f_0<1.711$, the stiffness is positive. Its value does not change significantly until $M_s$ starts to decrease. This happens exactly when the stiffness becomes negative, hence when the packing is unstable. Shortly thereafter the stiffness becomes positive again, while $M_s$ still continues to decrease.
The kinetic energy rises rapidly when $k<0$.  But at $f/f_0 =1.7111$, $k$ suddenly jumps to a positive value, and $E_{\mathrm{kin}}$ starts a rapid decline. This shows that the E$_\mathrm{kin}$ is controlled by $k$.
We showed this dependence in an earlier paper on failure in small packings \cite{Welker09}.
\begin{figure}
\centering
\includegraphics[angle=270, width=\columnwidth]{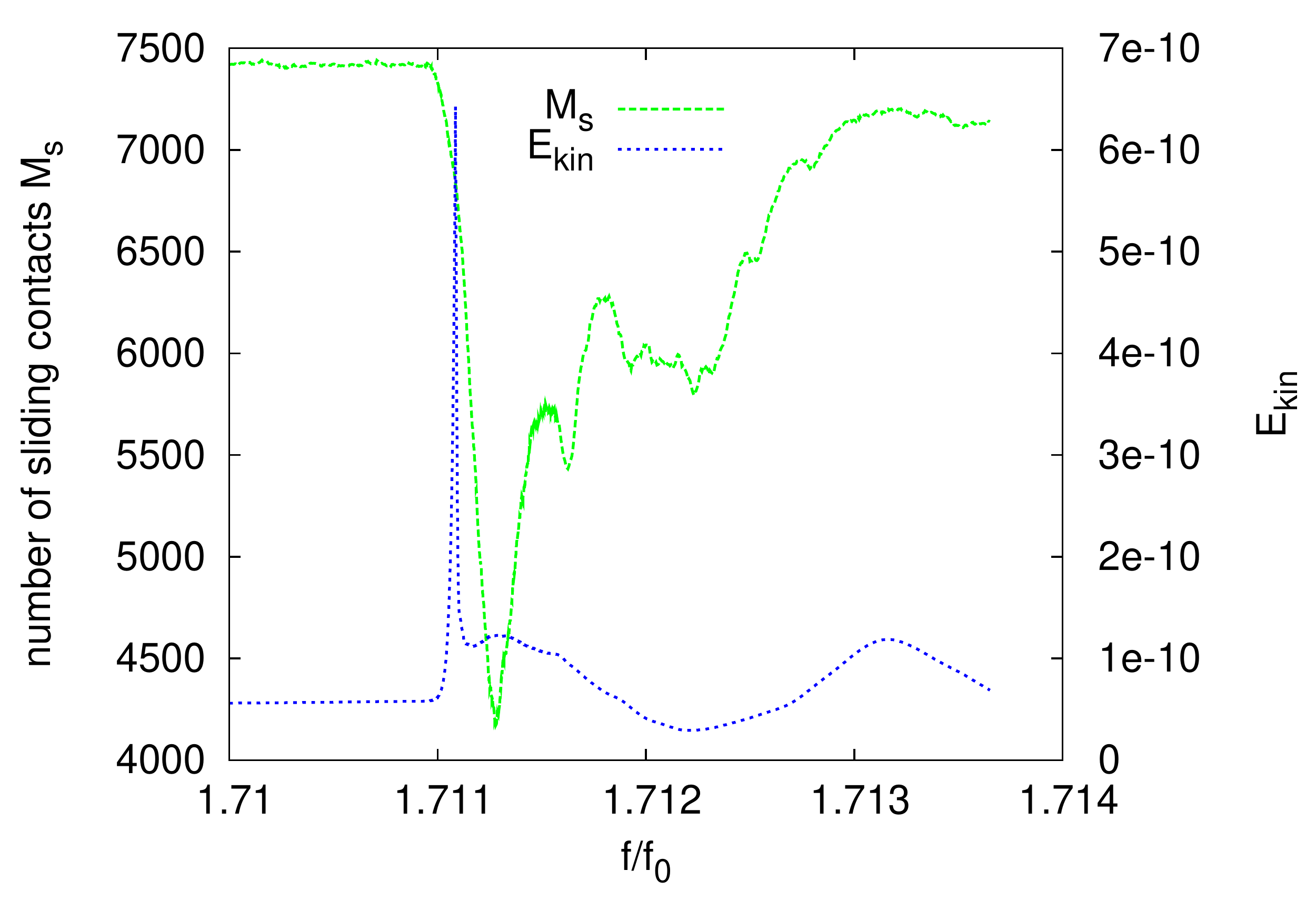}
\caption{Kinetic energy at the first precursors indicated in Figs~\ref{fig:stress_strain},~\ref{fig:KE},~\ref{fig:sliding},~\ref{fig:ms_changes} and~\ref{fig:ttest}.
The energy increases quickly, and then decreases again.}
\label{fig:KEpre}
\end{figure}

After the precursor, large vibrations appear that last for a long time
compared to the intervals shown in Figs \ref{fig:vkvpre},\,\ref{fig:KEpre}\,and\ref{fig:ttestprec}.
These vibrations can be observed in Fig.~\ref{fig:KE}.
We anticipated in \cite{Welker09} that vibrations will become important in large packings around failure.
The vibrations triggered by the precursor studied here do not cause failure, but Fig.~\ref{fig:KE} shows that vibrations grow as failure is approached. These vibrations may play an essential role in causing the collapse of the assembly.
Therefore failure might finally be initiated by the vibration generated by the precursors immediately preceding failure.

\subsubsection{Localization of the kinetic energy}
\label{sec:precvel}

\begin{figure}
\centering
\includegraphics[width=\columnwidth]{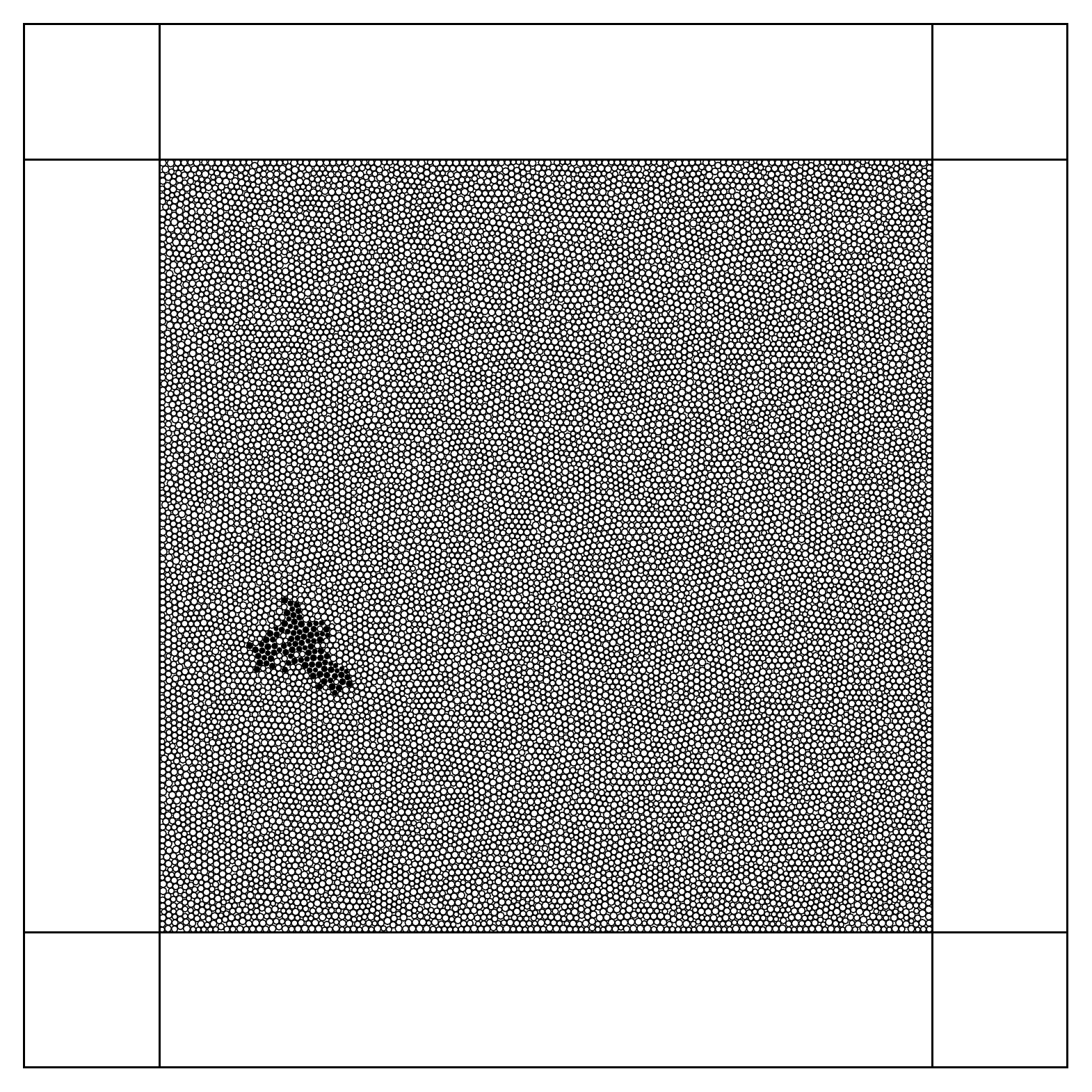}
\caption{Velocities in the assembly near the peak in $E_\mathrm{kin}$ (the last point of instability at $f/f_0=1.71108$ in Fig.~\ref{fig:vkvpre}).
The black particles (0.63\% of the total) have above average kinetic energy and carry 85\% of the total energy.
}
\label{fig:KE1}
\end{figure}

\begin{figure}
\centering
\includegraphics[width=\columnwidth]{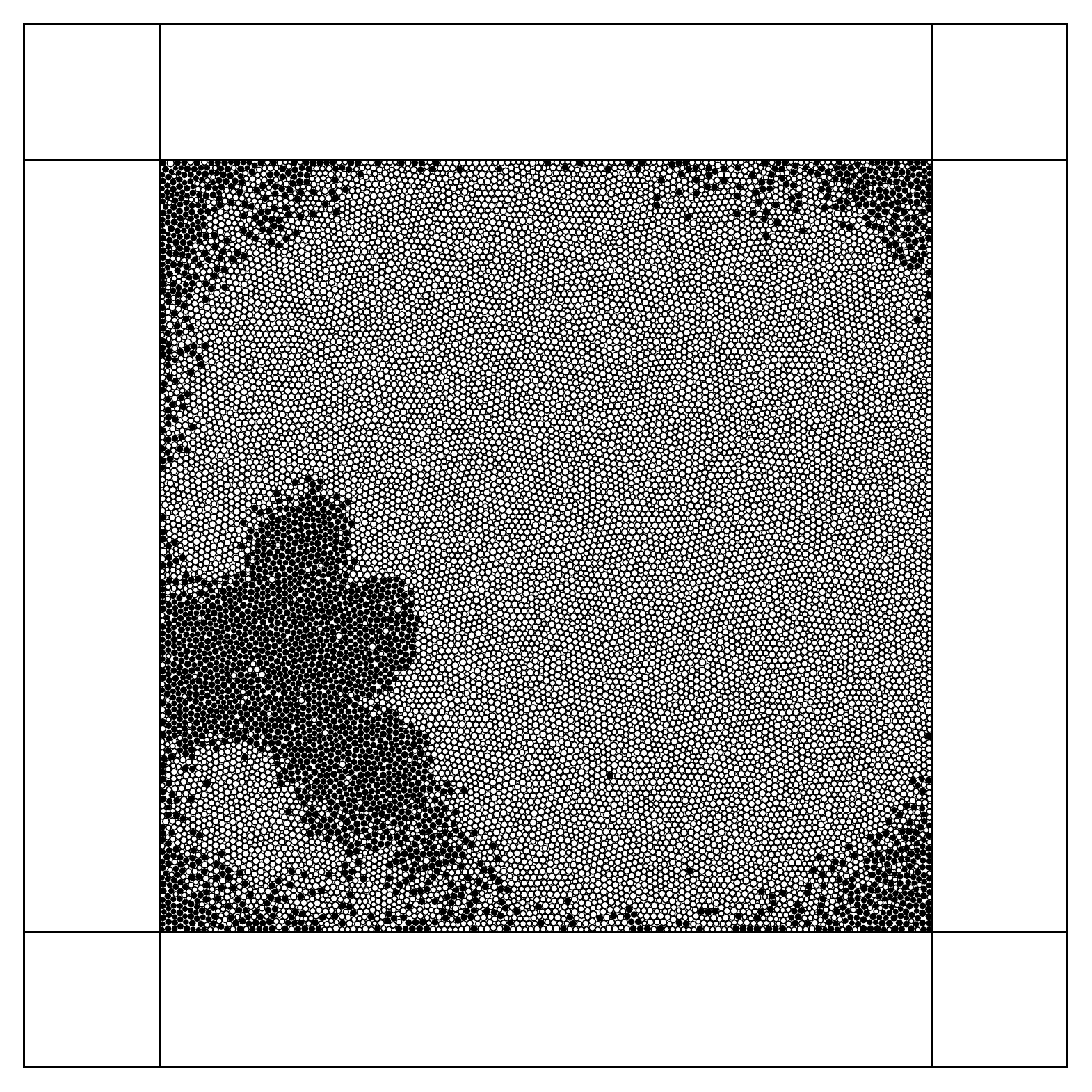}
\caption{Velocities in the assembly at $f/f_0=1.71112$, shortly after the recovery of stability in Fig.~\ref{fig:vkvpre}.
The black particles (16\% of the total) have above average kinetic energy and carry 64\% of the total energy.
}
\label{fig:KE2}
\end{figure}

When $E_\mathrm{kin}$ starts to rise at the beginning of the precursor, the velocities in a small region rise and become significantly larger than everywhere else.
Figure~\ref{fig:KE1} shows the grains of the packing that carry most of the kinetic energy near the peak of $E_\mathrm{kin}$.
But the instability for the precursor examined in Sec.~\ref{sec:prec:examine} lasts only for a short while, and the velocities in this region decrease again very soon. However, there is a wave of large movements spreading from this small region across the packing.
Figure~\ref{fig:KE2} shows the grains with large kinetic energy shortly thereafter, when the energy is propagated through the system.
In this energy spreading many different orientations are involved, and the propagated waves will move across the entire packing.
Looking at Figs~\ref{fig:vkvpre},~\ref{fig:KE1}, and Fig.~\ref{fig:KE2}, we see that the width of the spike in E$_\mathrm{kin}$ in Fig.~\ref{fig:vkvpre} is much smaller than the time it takes for the disturbance to cross the sample.
That means that the ``high'' energies ($> 10^{-10}$) occur only in a very localized region. This in turn shows that precursors are indeed ``localized failure''. 

\subsubsection{Where does the number of sliding contacts decrease?}
\label{sec:prec:msdecr}
\begin{figure}
\centering
\includegraphics[angle=270, width=\columnwidth]{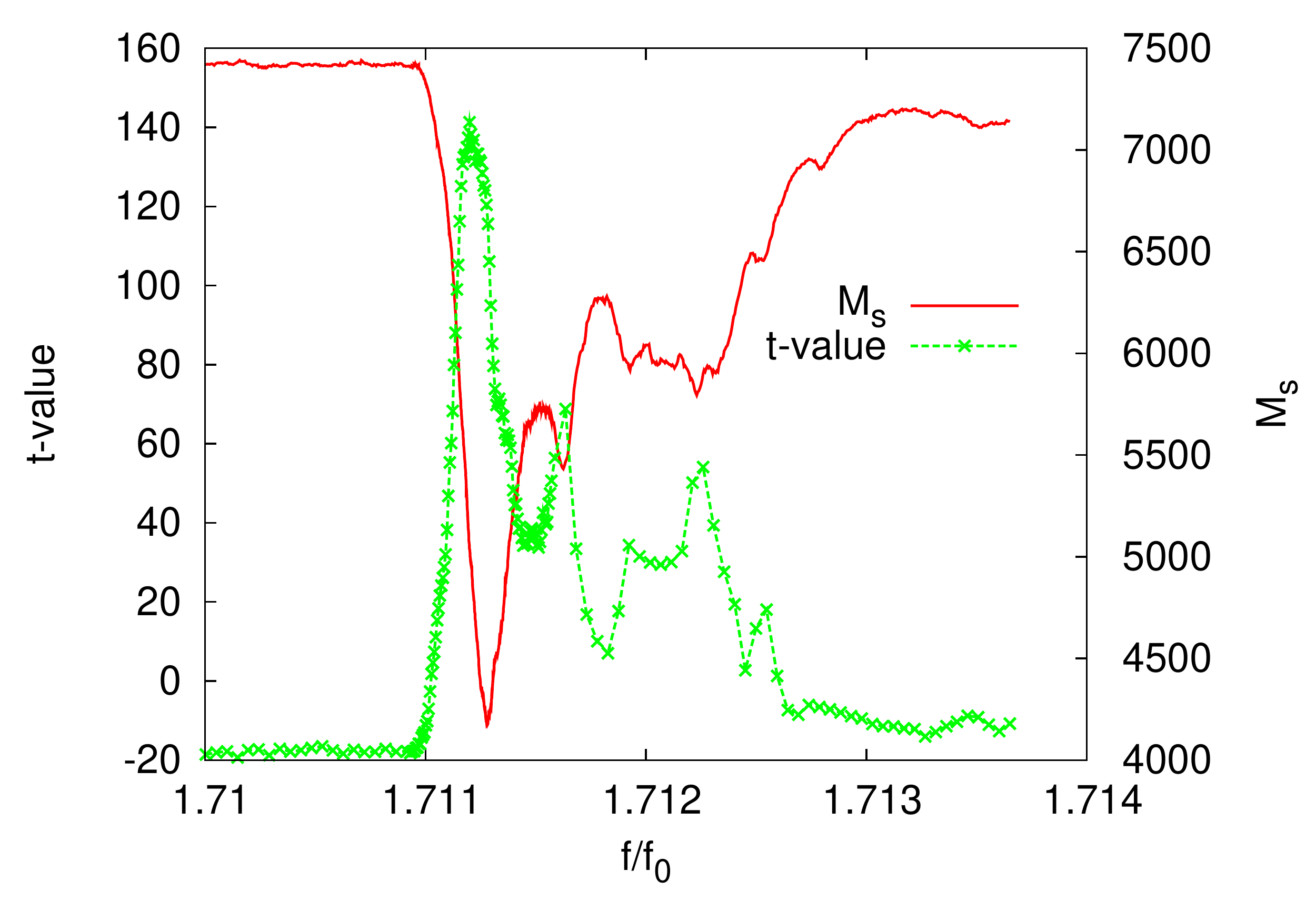}
\caption{Values from the $t$-test (Eq.~\ref{eq:ttest}) and $M_s$ during a precursor. The decrease in $M_s$ corresponds to a clustering of sliding contacts ($t>0$).}
\label{fig:ttestprec}
\end{figure}

\begin{figure}
\includegraphics[width=\columnwidth]{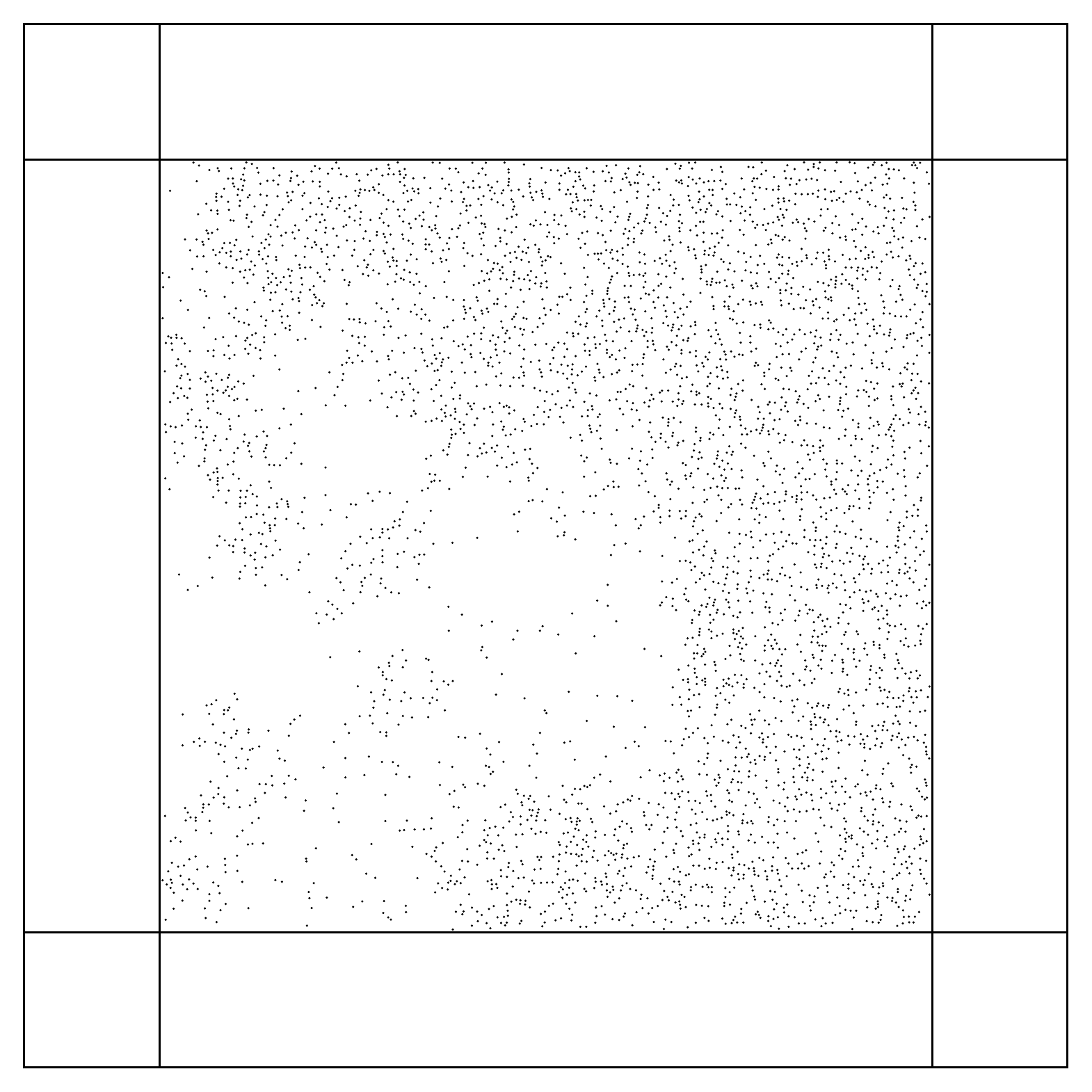}
\caption{Distribution of the number of sliding contacts in the packing at the time of the highest $t$-value in Fig.~\ref{fig:ttestprec}.}
\label{fig:sliding_prec}
\end{figure}

Figure~\ref{fig:ttestprec} shows $M_s$ and the $t$-test at the precursor. Before the precursor, the values are negative and do not change with increasing external force. When the precursor appears, the $t$-values increase strongly and become positive. Note that the maximum positive value is much larger than the negative value before the precursor.
It is reached at the time of minimum $M_s$.
Fig.~\ref{fig:sliding_prec} shows the spatial distribution of the sliding contacts at the maximum $t$-value. The sliding contacts disappear in some regions around the precursor, while the number of sliding contacts looks much more uniform far away from the precursor.
After the precursor, the $t$-values decrease again and become close to the values before the precursor. This indicates that ordering effects appear at the precursor, and disappear again after the precursor.
Comparing Fig.~\ref{fig:KE2} and Fig.~\ref{fig:sliding_prec}, we see that sliding contacts disappear in regions of elevated kinetic energy. Thus we conclude that the drop and subsequent recovery of $M_s$ are due to wave radiating outwards from the local failure. After the wave passes, the sliding contacts reappear, explaining why both $M_s$ and the $t$-test return to their initial values.

We anticipated in our last paper \cite{Welker09} that these \textsl{local failures} will occur.
One of our findings in \cite{Welker09} was that the number of sliding contacts vanishes at the failure. This statement is now extended to precursors, therefore these can indeed be called local failures.

\subsubsection{Change in the number of contacts}
\label{sec:prec:ms_ch}
\begin{figure}
\centering
\includegraphics[angle=270,width=\columnwidth]{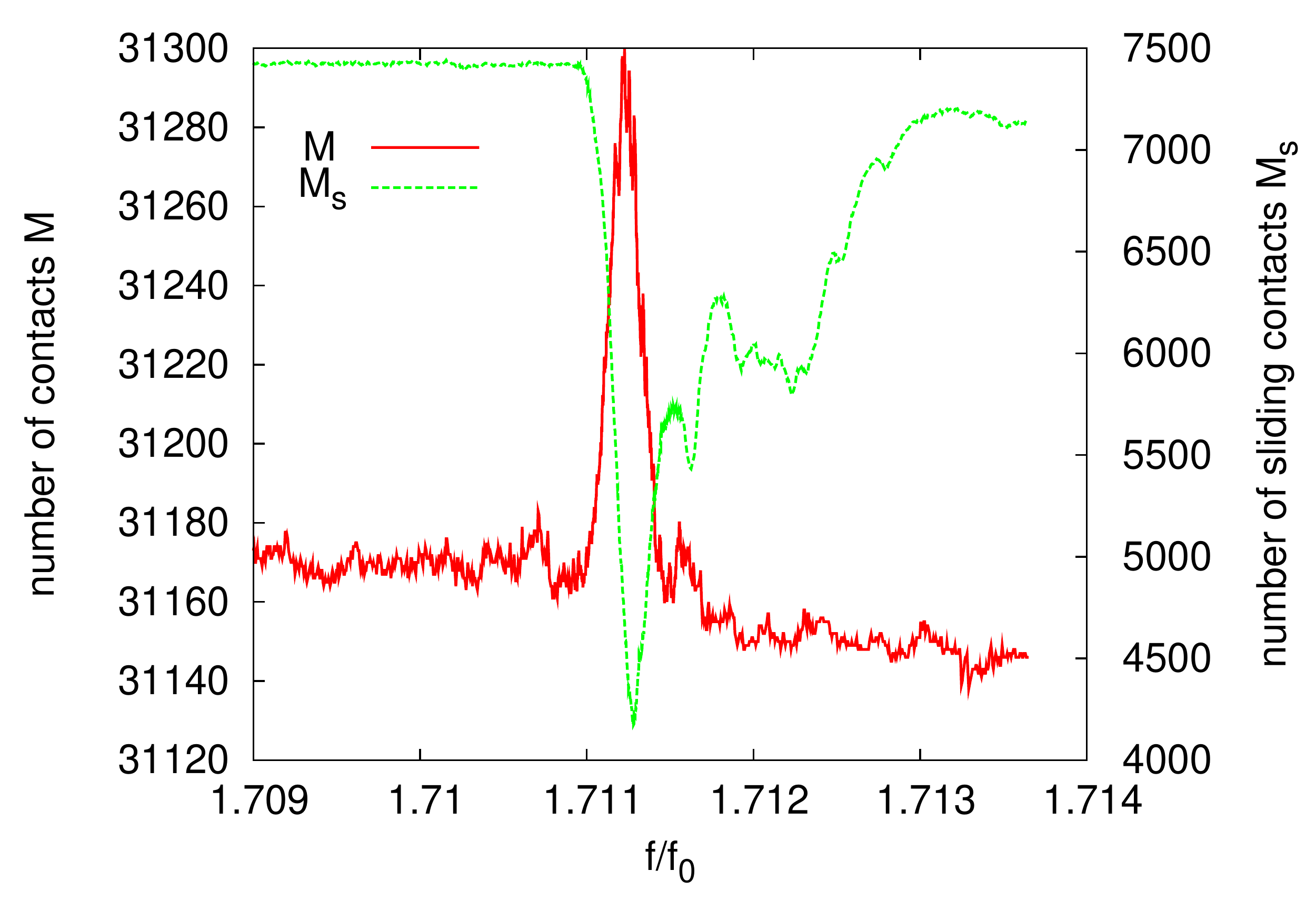}
\caption{Change in the number of contacts at one precursor.}
\label{fig:dM_precursor}
\end{figure}

Another feature of the precursor event is a transitional change in the total number of contacts. Fig.~\ref{fig:dM_precursor} shows
the total number of contacts $M$ as well as the number of sliding contacts $M_s$ during the precursor. Surprisingly, these quantities are anti-correlated: The sudden drop in the number of sliding contacts coincides with a peak in the number of contacts.  The contacts that are created are 
concentrated in the high kinetic energy regions of Figs~\ref{fig:KE1},~\ref{fig:KE2}. Therefore they are probably another effect of the wave.
When $M_s$ rises again, the number of contacts reduces slowly and attains values lower than those before the precursor.

\begin{figure}
\includegraphics[angle=270,width=\columnwidth]{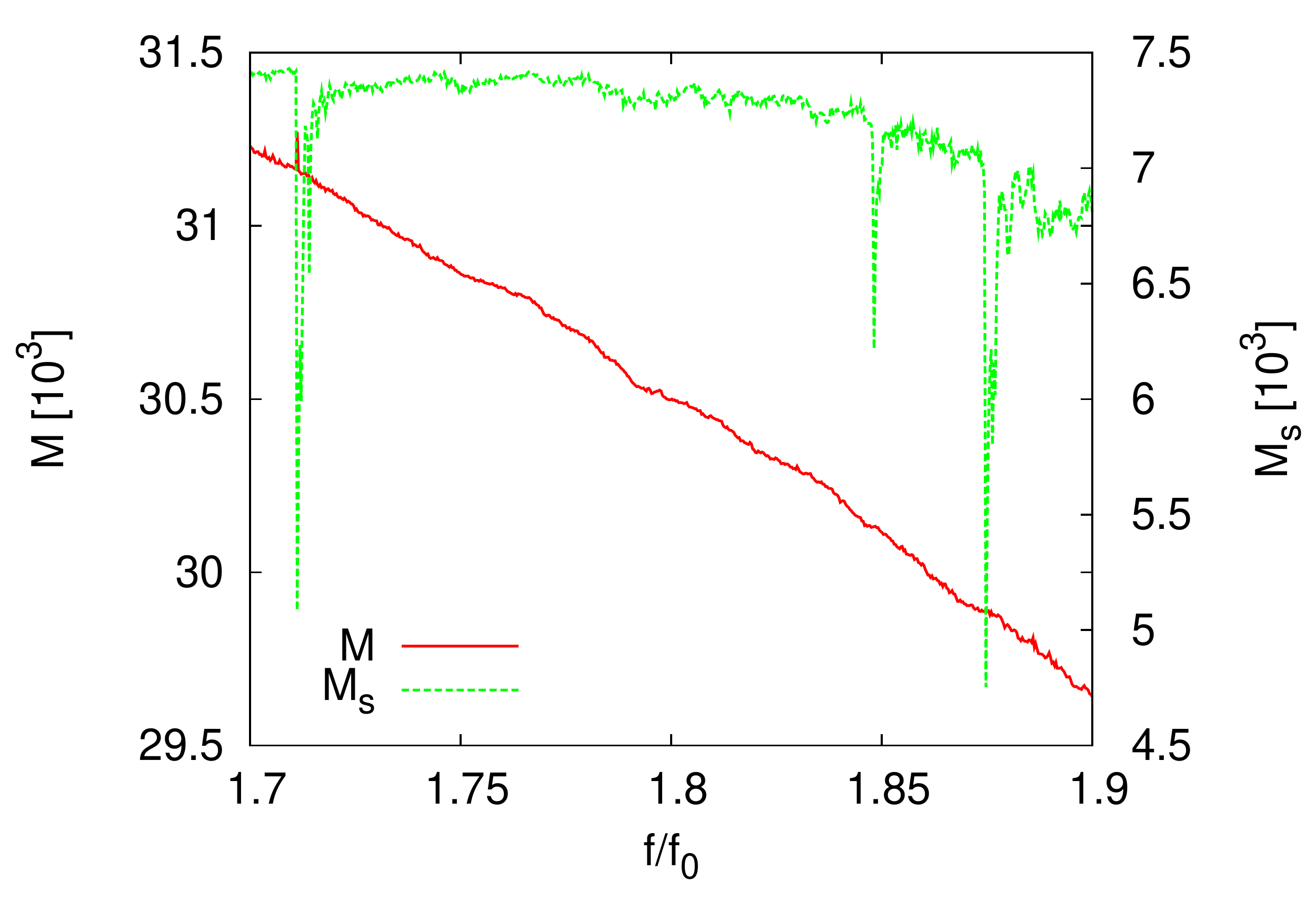}
\caption{Number of contacts in the assembly during a period where three precursors appear. The loss in the number of contacts seems not to be connected to the appearance of precursors.}
\label{fig:M}
\end{figure}

Another issue raised by Fig.~\ref{fig:dM_precursor} are the permanent changes induced in the contact network by the precursor.  For example, M decreases by about 30 between $f/f_0 = 1.709$ and $1.713$ -- is the precursor responsible for this change?  Plotting $M$ and $M_s$ over a long time (Fig.~\ref{fig:M}) shows that these changes are just part of a long term general trend.
Furthermore, if one identifies the contacts that have disappeared, one finds that they are not concentrated anywhere in particular.

We conclude therefore that the precursor does not lead to a significant change in the number of contacts.
Furthermore the geometric structure of the contact network remains almost unchanged.
Temporary changes mainly originate from an increase of the normal force $F_N$ at some contacts. This increase is generated by the compression wave radiated outward from the precursor.

\subsection{Can precursors be observed experimentally?}
\label{sec:prec:exp}
\begin{figure}
\centering
\includegraphics[angle=270, width=\columnwidth]{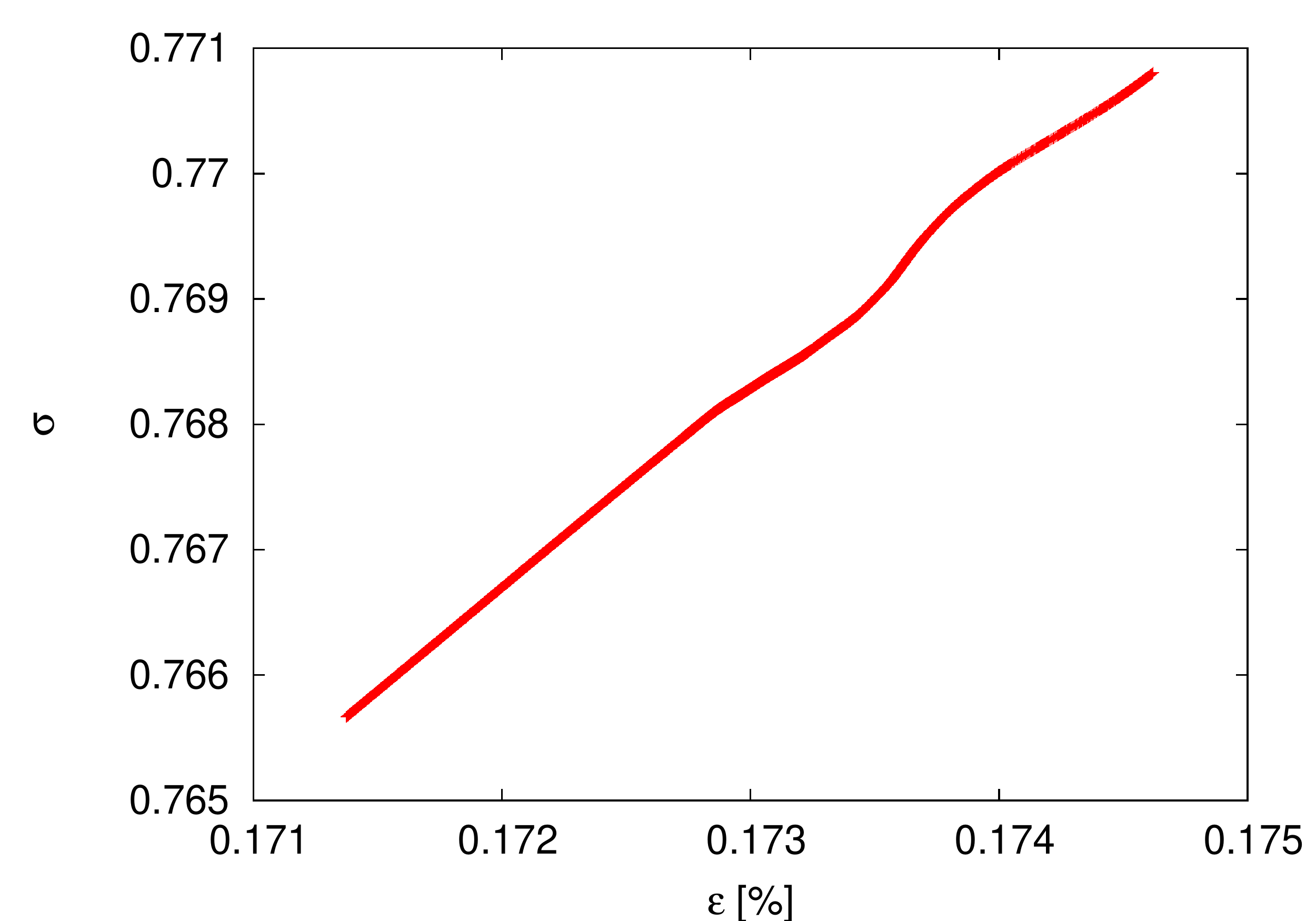}
\caption{Stress-strain relation during a precursor.
The range of the data corresponds to the range in Fig.~\ref{fig:KEpre}.
At the precursor the relation is not linear any more, but there is no visible peak at that time.}
\label{fig:stress_strain_precursor}
\end{figure}

While the stiffness of the assembly becomes negative and the kinetic energy rises, the stress-strain curve does not show a maximum at that time but rather a dip (Fig.~\ref{fig:stress_strain_precursor}). Thus in the stress-strain curve in Fig.~\ref{fig:stress_strain} it is hard to identify the precursor. Therefore it seems to be hard to even \textsl{notice} precursors in experimental biaxial (or triaxial) test, as it is difficult to detect small fluctuations of the stress-strain curve.

One possibility to detect precursors in experimental investigations is therefore to monitor the kinetic energy
by detecting sound emissions from these regions. These sound emissions 
arise at the local grain displacements \cite{Hidalgo02}.
Sound waves of high frequency are quickly diffused \cite{Jia04}, while low frequency waves can travel the packing almost unchanged
and can then be detected at the boundaries of the packing. By measuring the travel distance to different detectors, the spatial origin of the sound waves can be reconstructed \cite{Guarino98}.
However, one must take into account the dependence of the speed of sound on both the surface structure of the grains and the dimensionality (2D or 3D) \cite{Shourbagy08}.
Sound waves have been observed in triaxial tests but not analyzed so much.

\subsection{Summary}
\label{sec:prec:summary}
The definition of precursors is based on the observation that the number of sliding contacts decreases suddenly at certain values of the external force $f$. A closer inspection of one precursor shows that there is a multifaceted behavior at this time: first of all, the precursor is initiated by an instability. Through this instability, the kinetic energy rises quickly, but only in a limited region of the packing. The rising energy initiates the observed decrease in the number of sliding contacts. The further decrease in $M_s$ is then governed by wave radiation.
This wave also increases the total number of contacts.
Both of these changes are transitional. When the wave is gone, $M$ and $M_s$ almost return to their values before the precursor.
A closer inspection of the contact status changes furthermore reveals that only a few contacts permanently disappear (38) or are created (14).

We therefore conclude that the precursors are localized instabilities, in contrast to failure, i.e., the global loss of stability.

\section{Conclusion}
\label{sec:conclusion}
We investigated the behavior of granular packings in two dimensions that are submitted to an increasing external force along one direction.
We performed this biaxial test under quasi-static conditions where the forcing is slowly increased to a critical force. This critical force leads to a large deformation of the packing.

We found that the time leading up to failure can be divided into two roughly equal periods. Many variables show a qualitatively different behavior in these two periods.
For example,
the volume decreases in the first period, but increases in the second.
The number of sliding contacts increases linearly in the first period, and then decreases in the second, attaining a maximum near the transition between the two behaviors.
Thus the number of sliding contacts
is not related univocally to the stiffness or the stability of the packing.
The spatial organization of sliding contacts also changes during the simulation.
Before the maximum, sliding contacts are more uniformly distributed in the packing than afterwards.
Comparison with a Poisson process shows that sliding contacts initially repel each other: the presence of a sliding contact reduces the probability that a neighboring contact will become sliding.
After the maximum, in the second regime, the situation is reversed: sliding contacts are concentrated in specific regions.
Near failure, the formation of the shear band is foreshadowed by a concentration of sliding contacts. These contacts cluster preferably to form diagonal bands,
This suggests that the localization of deformation begins long before any shear band is visible.

The changes in the number of sliding contacts are caused by the frequency of the different contact statuses transitions; in the first period, the main transition is from closed to sliding, leading to an increase in $M_s$. In the second period, the transition sliding to open is dominating, decreasing $M_s$.

Around the time of the maximum number of sliding contacts, near the transition between the two regimes, precursors begin to appear, becoming more and more frequent as failure is approached.
These precursors are triggered by instabilities that lead to a sudden rearrangement of a small, localized number of grains who carry most of the kinetic energy. Precursors involve also a strong decrease in the number of sliding contacts, and a temporary increase in the number of contacts. When stability is recovered, the packing relaxes to a new equilibrium with properties ($M$, $M_s$, $E_\mathrm{kin}$, $\ldots$) close to the values before the precursor.

Precursors are initiated by instabilities, therefore inertia effects become important. Hence, when investigating the micro-macro transition, inertia cannot be neglected any more, complicating the establishment of a macroscopic theory based on microscopic, static quantities such as the fabric tensor or the number of sliding contacts.

The appearing instability during the precursor leads to large vibrations involving motions of all particles.
These motions let the packing explore a larger part of the phase space.
When approaching the critical external force, the packing becomes very soft, and the vibrations, triggered through the precursors, become larger. We therefore argue that the precursors observed in this study should be significantly involved in failure.
This supposition might be investigated in a future article.
\par
Financial support of the DFG through SFB716, project B3, is acknowledged.

\appendix

\section{The stiffness matrix}
\label{app:vkv}
In Sec.~\ref{sec:prec:instab},
we used the quantity $k=\mathbf{vkv}/\mathbf{vv}$ to estimate the stiffness
of the packing.
This stiffness is the sum of a mechanical part $k_\mathrm{mech}$ and a geometric contribution $k_\mathrm{geo}$. Usually the mechanical part is dominating, $k_\mathrm{mech}\gg k_\mathrm{geo}$.
In Ref.~\cite{Welker09}, it has been shown that this quantity is connected
to the macroscopic stiffness of the packing, i.e., it shows how large is the deformation for a certain change in load (Note that in \cite{Welker09} $k$ is defined to be $k_\mathrm{mech}$).

The stiffness matrix has been extensively discussed in Refs.~\cite{Welker09,McNamara06}.
It arises when one writes the equations of motion for all the particles in vector form.
More specifically, one
forms the vector $\mathbf{v}$ containing the velocities (translational
and angular) of all the particles.
In a two-dimensional system, it has $3N$ components,
where $N$ is the number of particles.
If the motion is quasi-static, then $-\mathbf{kv}$ is the temporal
derivative of the contact forces exerted on each particle.
When the packing is stable, these forces balance the applied load.

One criteria for stability is $\mathbf{vkv}>0$, i.e. positive stiffness.
(We normalize $\mathbf{vkv}$ by $\mathbf{vv}$ so that fluctuations in velocity
do not affect $k$.
Note that $-\mathbf{vkv}/\mathbf{vv}$ is the stiffness' contribution to the second derivative of the kinetic energy $E_\mathrm{kin}$.)
In small systems, failure often occurs when a contact status change
leads to a modification of $\mathbf{k}$ that makes $k<0$ \cite{Welker09}.
Now $\mathbf{vkv}<0$ implies that (the symmetric part of) $\mathbf{k}$
has at least one negative eigenvalue.
The amplitude of its eigenvector grows exponentially during the instability.

In large packings, the rise in $E_\mathrm{kin}$ is localized, therefore only a small number of velocities control the change of $k$. More specifically, at least one of the eigenvalues of (the symmetric part of) $\mathbf{k}$ must be negative at the precursor, and its eigenvector $\mathbf{v}_\mathrm{*}$ defines which particle velocities grow quickly with time. This growing velocities consequently define the stiffness ($k\approx \mathbf{v}_\mathrm{*}\mathbf{k}\mathbf{v}_\mathrm{*}$/$\mathbf{v}_\mathrm{*}\mathbf{v}_\mathrm{*}$).
This also shows why the jump in $k$ from negative to positive \textsl{must} appear exactly at the maximum in $E_\mathrm{kin}$: at that time, the negative eigenvalue becomes positive. Therefore $k$ jumps, as the corresponding eigenvector $\mathbf{v}_\mathrm{*}$ is large.
Note that if $\mathbf{v}_\mathrm{*}$ did not grow so much, it would not control $k$. Therefore it is probable that small precursors with a small loss of sliding contacts do not lead to a negative value of $k$.

Remark: the radiation of the wave is a dynamic process, and therefore not captured by $d\mathbf{f}_\mathrm{ext}/dt = \mathbf{kv}$.

\end{document}